\documentclass[sigconf,screen]{acmart}

\setcopyright{none}
\renewcommand\footnotetextcopyrightpermission[1]{} 
\usepackage{graphicx}
\graphicspath{{images/}}
\usepackage{subcaption}
\usepackage{paralist}
\usepackage{enumitem}
\usepackage{multirow}
\usepackage{makecell}
\usepackage{url}

\AtBeginDocument{%
  }

\newif\ifcomment
\commentfalse
\definecolor{darkblue}{RGB}{0, 0, 160}

\ifcomment
	\newcommand{\edc}[1]{\textbf{\color{darkblue}***EDC: #1***}}
	\newcommand{\sundar}[1]{\textbf{\color{blue}MS: #1}}
  \newcommand{\igor}[1]{\textbf{\color{violet}Igor: #1}}
  \newcommand{\final}[1]{{\color{red}#1}}
\else
	\newcommand\edc[1]{}
	\newcommand\sundar[1]{}
  \newcommand\igor[1]{}
  \newcommand\final[1]{#1}
\fi

\newif
\iflong
\longtrue
\iflong
\newcommand{\shortVer}[1]{}
\newcommand{\longVer}[1]{#1}
\else
\newcommand{\shortVer}[1]{#1}
\newcommand{\longVer}[1]{}
\fi

\newcommand{\descr}[1]{\smallskip\noindent\textbf{#1}}
\newcommand{\sspace}{\vspace{10pt}}
\newcommand{\nspace}{\vspace{-6pt}}

\begin{document}

\title[Beyond the Crawl: Unmasking Browser Fingerprinting in Real User Interactions]{Beyond the Crawl: Unmasking Browser Fingerprinting in\\Real User Interactions}
\titlenote{A slightly shorter version of this paper appears in the Proceedings of the 34th ``The Web Conference'' (WWW 2025). Please cite the WWW version.}

\author{Meenatchi Sundaram Muthu Selva Annamalai}
\email{meenatchi.annamalai.22@ucl.ac.uk}
\affiliation{%
  \institution{University College London}
  \country{United Kingdom}
}

\author{Emiliano De Cristofaro}
\email{emilianodc@cs.ucr.edu}
\affiliation{%
  \institution{University of California, Riverside}
    \country{United States\\\em Corresponding Author}
}

\author{Igor Bilogrevic}
\email{ibilogrevic@google.com}
\affiliation{%
  \institution{Google, Zurich}
	  \country{Switzerland}
}

\begin{abstract}
Browser fingerprinting is a pervasive online tracking technique used increasingly often for profiling and targeted advertising.
Prior research on the prevalence of fingerprinting heavily relied on automated web crawls, which inherently struggle to replicate the nuances of human-computer interactions.
This raises concerns about the accuracy of current understandings of real-world fingerprinting deployments.
As a result, this paper presents a user study involving 30 participants over 10 weeks, capturing telemetry data from real browsing sessions across 3,000 top-ranked websites.

Our evaluation reveals that automated crawls miss almost half (\final{45\%}) of the fingerprinting websites encountered by real users.
This discrepancy mainly stems from the crawlers' inability to access authentication-protected pages, circumvent bot detection, and trigger fingerprinting scripts activated by specific user interactions.
We also identify potential new fingerprinting vectors present in real user data but absent from automated crawls.
Finally, we evaluate the effectiveness of federated learning for training browser fingerprinting detection models on real user data, yielding improved performance than models trained solely on automated crawl data.
\end{abstract}

\maketitle
\pagestyle{plain}

\section{Introduction}

Browser fingerprinting is an invasive online tracking technique widely considered to be a significant threat to users' privacy~\cite{firefoxbrowserfp2020,w3cbrowserfp,applefp}.
Generally, it involves trackers using client-side device information (e.g., hardware specifications or browser configurations) to derive unique device identifiers and track users across multiple visits and websites.
Unlike third-party cookies, browser fingerprinting is \textit{stateless} and thus less visible to users, who have limited control and few tools to defend against it.

Worse yet, browser fingerprinting is highly intrusive, as the identifiers remain stable over long periods of time~\cite{pugliese2020long} and can be effective even when using incognito mode~\cite{akhavani2021browserprint}.
Countermeasures can be tricky to deploy after detection, as they often lead to significant website breakage and affect user experience~\cite{iqbal2021fingerprinting,mozillabreakage}, not to mention that the unique identifier may have already been revealed at that point.
With the increasing prevalence of fingerprinting~\cite{iqbal2021fingerprinting} and the phasing out of third-party cookies by major browsers, countering invasive browser fingerprinting has become crucial to protecting user privacy on the modern Web.

Detecting fingerprinting attempts, often implemented through JavaScript scripts, is a prerequisite for effective mitigation.
Initially, these scripts were mainly identified through simple heuristics~\cite{englehardt2016online,acar2014web} and manually curated blocklists~\cite{disconnect2018,easyprivacy,privacybadger}; however, these are hard to maintain and often narrowly defined to reduce false positives, thus missing many fingerprinting scripts in practice.
Consequently, machine learning (ML) based detectors have become increasingly popular~\cite{das2018web,ikram2017towards,iqbal2021fingerprinting} and can detect significantly more fingerprinting scripts with comparable false positive rates to heuristics. %
Overall, these techniques generally rely on centralized, automated crawlers instructed to visit a large number of websites %
to collect scripts for analysis and fingerprinting detection.

One major limitation of automated crawlers is their limited ability to faithfully replicate genuine user behaviors and interactions.
Although recent work has made progress toward emulating some degree of user interactions (such as accepting cookie banners~\cite{senol2024double,liu2024identified} and emulating user interests in browsing patterns~\cite{liu2024identified}), many key actions remain firmly outside of the reach of automated crawlers.
For instance, these often fail to reliably solve CAPTCHAs, evade bot detectors, login, access websites behind paywalls, etc.
This leads to incomplete website coverage and potentially missed fingerprinting scripts.
For example,~\citet{iqbal2021fingerprinting}'s automated crawl failed to visit 11.9\% of top-ranked websites.

To address this limitation, prior work has investigated the feasibility of training ML models on real-user browsing sessions.
Evidently, this needs to be done in a privacy-preserving way to avoid exposing users' browsing histories, etc.
\citet{annamalai2024fp} recently proposed FP-Fed, a system relying on differentially private federated learning (DP-FL) to collaboratively train ML models on the combined browsing sessions of many users while providing strong formal privacy guarantees. %
Although FP-Fed can train models that achieve good performance even at moderate levels of privacy, the authors only tested it on a dataset derived from an automated crawl, with unclear implications for its performance in the real world. 
\descr{Technical Roadmap.}
In this paper, we investigate the prevalence and distribution of browser fingerprinting in real-user browsing sessions, as opposed to automated crawls used in prior work~\cite{englehardt2016online,iqbal2021fingerprinting,annamalai2024fp}.
Specifically, we collect real-world data from a 10-week study (June-August 2024) involving 30 participants who browse the Web with their own devices and report telemetry via a Chrome extension we provide.
This enables us to collect browser fingerprinting signals from real browsing sessions across 3,000 top-ranked websites.

The resulting analysis allows us to shed light on two distinct aspects of fingerprinting.
First, we compare the prevalence and distribution of browser fingerprinting in a real-world dataset to an equivalent automated crawl of the same 3,000 websites.
Second, we conduct a comparative analysis of the ML performance of browser fingerprinting detectors trained on the automated crawl and the real-world dataset in a distributed and privacy-preserving way.

\descr{Main Findings.} Overall, our study and experimental analysis yield three interesting findings:
\begin{itemize}[leftmargin=15pt]
\item Surprisingly, {\em almost half } (\final{45\%}) of fingerprinting websites identified from real-user browsing sessions are missed by automated crawls.
This is due to three main reasons: (1) authentication pages that \final{typically require specific user interactions (e.g., logging in)}, (2) bot detection scripts that track real user interaction before browser fingerprinting is triggered, and (3) cookie banners that require user consent -- \final{before fingerprinting scripts are triggered.} \smallskip

\item We discover potential new fingerprinting vectors from the real-user browsing sessions that were not previously found in the automated crawls. \smallskip

\item We show that ML models trained privately on real-user browsing sessions can achieve comparable or even better performance than non-private models trained on automated crawls alone while providing privacy and learning the behavior of many more fingerprinting scripts.
Specifically, in our experiments, the former achieves an Area Under the Precision-Recall Curve (AUPRC) of \final{0.98} at a privacy level of $\varepsilon = 5$, compared to an AUPRC of \final{0.96} for the latter.
\end{itemize}

Overall, our work paves the way for \final{future} deployment of more effective, dynamic, and robust browser fingerprinting detection relying on a scalable, distributed, and privacy-preserving infrastructure geared to be readily integrated into modern browsers.

\section{Background \& Related Work}
In this section, we review browser fingerprinting and differentially private federated learning, along with relevant prior work.

\subsection{Browser Fingerprinting}
\label{sec:browser_fp}
Browser fingerprinting is a tracking technique usually deployed through Javascript running on a user's browser to build a unique user identifier.
This user identifier typically consists of high-entropy device information (e.g., screen size, GPU model) used to produce highly unique and stable user identifiers.
Unlike third-party cookies, browser fingerprinting is \textit{stateless}, i.e., no data is stored on the user's device and, therefore, cannot be easily detected or mitigated.
Although browser fingerprinting has been known to be used for legitimate purposes, e.g., Web authentication~\cite{alaca2016device,laperdrix2019morellian,senol2024double} or fraud detection~\cite{iovationfraud2019,relix2018}, it is often used for online tracking and to serve targeted ads~\cite{liu2024identified}.
As a result, browser fingerprinting is widely considered a significant threat to user privacy~\cite{w3cbrowserfp}, thus prompting many browser vendors to deploy countermeasures~\cite{firefoxbrowserfp2020,applefp,bravefp}.

Although there are well-documented cases of fingerprinting in the wild, we are not aware of a widely accepted formal definition of fingerprinting~\cite{iqbal2021fingerprinting}.
Mayer~\cite{mayer2009any} was the first to observe that the uniqueness and customization of browsing environments (which they called ``quirkiness'') can be abused to identify users.
The large-scale Panopticlick experiment later conducted by the late Peter Eckersley~\cite{eckersley2010unique} showed that most browsing sessions (83.6\%) have unique fingerprints.
While these results are concerning on their own, the fingerprinting surface considered by this early work was relatively limited to information collected from simple Javascript APIs like the \texttt{Screen} and \texttt{Navigator} APIs and HTTP headers.
More recently, as more features and APIs were added to the Javascript specification, fingerprinting has begun to include Canvas~\cite{mowery2012pixel}, Audio~\cite{englehardt2016online}, WebRTC~\cite{englehardt2016online}, WebGL~\cite{cao2017cross}, Battery Status~\cite{olejnik2016leaking}, Mobile Sensor~\cite{bojinov2014mobile}, and even the Web Bluetooth APIs~\cite{bahrami2021}.
With such a large fingerprinting surface, it is often difficult to pinpoint the \textit{intent} behind the use of these various APIs in arbitrary Javascript scripts, thus making it challenging to specify a single comprehensive definition of fingerprinting.

Nevertheless, we follow prior work by Iqbal et al.~\cite{iqbal2021fingerprinting} and take a conservative approach to defining fingerprinting based on a well-known high precision (low false positive rate) heuristic.
Specifically, we exclude the simple curation of properties from the \texttt{Navigator} and \texttt{Screen} APIs and focus on the four most obvious forms of fingerprinting (Canvas, Canvas Font, Audio, and WebRTC).
This definition not only minimizes the probability of flagging non-fingerprinting scripts but is also helpful in training ML models that generalize well to other forms of fingerprinting as well~\cite{annamalai2024fp}.

\smallskip Next, we briefly describe the four main types of fingerprinting identified by the heuristics~\cite{englehardt2016online} and their associated detection criteria.  

\descr{\em Canvas.} Scripts exploit differences in how fonts are rendered across devices.
Criteria:
\begin{compactenum}
  \item text is written to the \texttt{canvas} element using the \texttt{fillText} or \texttt{strokeText} method;
  \item style is applied with \texttt{fillStyle} or \texttt{strokeStyle} method;
  \item \texttt{toDataURL} is called to extract the image from the \texttt{canvas}; and
  \item \texttt{save}, \texttt{restore} or \texttt{addEventListener} methods are not called.
\end{compactenum}

\descr{\em Canvas Font.} Scripts access the list of fonts installed on a device.
Criteria:
\begin{compactenum}
  \item \texttt{font} property of canvas element is set to more than 20 different fonts; and
  \item \texttt{measureText} is called more than 20 times.
\end{compactenum}

\descr{\em WebRTC.} Scripts rely on the uniqueness of peers in the WebRTC protocol~\cite{webrtc}.
Criteria:
\begin{compactenum}
  \item \texttt{createDataChannel} or \texttt{createOffer} method is called on a WebRTC peer connection; and
  \item \texttt{onicecandidate} or \texttt{localDescription} method is called.
\end{compactenum}

\descr{\em AudioContext.} Scripts exploit differences in the different hardware processing audio.
The detection criterion is that at least one of
 \texttt{createOscillator}, \texttt{createDynamicsCompressor}, \texttt{destination}, \texttt{startRendering}, or \texttt{oncomplete} are called.

\subsection{Detecting Browser Fingerprinting}
Early research on browser fingerprinting detection has mainly relied on manually curated blocklists, e.g., EasyPrivacy~\cite{easyprivacy}, Privacy Badger~\cite{privacybadger}, and Disconnect~\cite{disconnect2018}.
However, as fingerprinting vendors and scripts constantly evolve, it can be challenging to maintain these lists continuously.

\descr{Heuristics-based detection.}
Manual analysis of fingerprinting scripts by~\citet{acar2013fpdetective} and~\citet{roesner2012detecting} paved the way for the first heuristic that detected canvas fingerprinting automatically~\cite{acar2014web}.
More precisely,~\citet{acar2013fpdetective} monitored and analyzed the arguments and return values of the \texttt{fillText}, \texttt{strokeText}, and \texttt{toDataURL} methods exposed by the \texttt{Canvas} API.
Three additional types of fingerprinting (Canvas Font, WebRTC, and Audio) were later added by~\citet{englehardt2016online}.
This combined heuristic is now widely used as a prominent indicator of fingerprinting~\cite{das2018web,iqbal2021fingerprinting,annamalai2024fp,senol2024double} as it is known to produce very few false positives, which is an important consideration to prevent falsely flagging fingerprinting scripts.
However, as noted by~\citet{iqbal2021fingerprinting}, it might miss many fingerprinting scripts in practice since it is defined very narrowly to achieve high precision.
Furthermore, keeping the heuristic up-to-date with the latest Javascript APIs and fingerprinting vectors can be challenging.

\descr{ML-based detection.}
Methods based on machine learning solve the problem of manually maintaining blocklists and heuristics by learning fingerprinting behaviors in the wild.
\citet{ikram2017towards} trained a one-class SVM on \textit{static} features directly extracted from scripts' source code.
However,~\citet{iqbal2021fingerprinting} observed that code is often obfuscated, making it difficult to reliably learn fingerprinting behaviors from static features alone; therefore, they additionally trained a Decision Tree classifier on \textit{dynamic} features extracted from the execution trace of a given script.
By monitoring and analyzing the number of times a script calls each Javascript API, along with the associated arguments and return values,~\citet{iqbal2021fingerprinting} trained a robust browser fingerprinting detector that achieves both high precision and high recall.

Recently,~\citet{annamalai2024fp} proposed going beyond \textit{centralized} models that rely on a large automated crawl, which cannot replicate human interaction and, therefore, might miss fingerprinting scripts in the wild. %
They proposed FPFed, a system geared to train a browser fingerprint detection model collaboratively on real users' browsing sessions while preserving privacy using the federated learning paradigm~\cite{mcmahan2016federated} (see below).
However,~\citet{annamalai2024fp} did not actually evaluate their system on real-world user behavior but opted to test their system by simulating real-world users on an automated crawl.
Our work aims to fill this research gap by collecting browsing data from real users and evaluating the FP-Fed system on a real-world dataset instead.

\subsection{Differentially Private Federated Learning}
The standard way to train models collaboratively on multiple users' data while providing formal privacy guarantees is through differentially private federated learning (DP-FL).
More precisely, DP-FL combines Federated Learning (FL) and Differential Privacy (DP). %

\descr{Federated Learning.}
Typically, centralized ML models are trained on datasets stored at a single entity.
However, in many scenarios, gathering data from multiple users may not be possible due to privacy, security, and/or efficiency concerns.
To this end, Federated Learning (FL)~\cite{mcmahan2016federated} introduces a distributed learning approach allowing users to train an ML model collaboratively without disclosing their (potentially sensitive) training data.

In FL, users train local models on their individual training data, sharing only their model updates (and not the raw training data) with a server.
The server aggregates the model updates to build a global model, which is then propagated back to the users.
This process then repeats until the global model converges.
Although FL ensures that the server never sees the raw training data, prior work has shown that the model updates can still leak sensitive information about users' training data~\cite{melis2019exploiting}.
Therefore, recent work has focused on providing formal privacy guarantees. %

\descr{Differential Privacy (DP).}
DP is the standard framework for defining algorithms that provide theoretical upper bounds on the loss of privacy incurred by data subjects due to the output of an algorithm~\cite{dwork2014algorithmic}. 

\begin{definition}[Differential Privacy (DP)]
  A randomized mechanism $\mathcal{M} : \mathcal{D} \rightarrow \mathcal{R}$ is $(\varepsilon, \delta)$-differentially private if for any two neighboring datasets $D, D' \in \mathcal{D}$ and $S \subseteq \mathcal{R}$: 
  \begin{equation*}
    \Pr[\mathcal{M}(D) \in S]  \leq e^\varepsilon \Pr[\mathcal{M}(D') \in S] + \delta 
  \end{equation*}
\end{definition}

The privacy guarantees provided by a differentially private algorithm are parameterized by $\varepsilon$ and $\delta$.
The privacy parameter, $\varepsilon$, is often referred to as the privacy budget and ranges from 0 to $\infty$, with lower values denoting better privacy.
Whereas $\delta$ quantifies the probability that the mechanism fails to deliver any guarantees; %
typically, $\delta$ is fixed to some asymptotically small number (e.g., $10^{-5}$).

\descr{DP-FL.}
DP can be combined with FL in many ways to provide strong, formal privacy guarantees in the FL setting.
One way is by adding statistical noise when the server aggregates the model updates, aka Central DP (CDP).
Under this regime, the model is trusted with the aggregated model updates but not the raw sensitive data.
CDP guarantees that it is impossible (up to the privacy parameter $\varepsilon$) to infer whether or not data from a user was used to train the global model based on the aggregated and noised model updates.
Examples of CDP instantiations for DP-FL include next-word prediction~\cite{mcmahan2018learning}, medical image analysis~\cite{adnan2022federated}, and network analysis~\cite{naseri2022cerberus}.
Other approaches use Local DP~\cite{truex2020ldp,sun2021ldp} or Distributed DP~\cite{kairouz2021distributed} to minimize trust assumptions in the server.

In this work, we consider the FP-Fed~\cite{annamalai2024fp} setting, which uses CDP, as it provides a good trade-off between utility, convergence speed, and the amount of noise required for the desired level of privacy.
A detailed overview of the FP-Fed can be found in Section~\ref{sec:fpfed}. %

\section{Methodology}

Next, we discuss our methodology to collect browser fingerprinting scripts from real users' browsing sessions.
We describe the websites we collect data from and how we recruit participants, collect website telemetry, and detect fingerprinting.
We also discuss relevant ethical considerations with respect to collecting and analyzing the data.

\subsection{Websites of Interest}
Ideally, to capture real user interactions, we would want to collect and analyze telemetry data from all websites visited by participants. %
However, browsing histories are highly sensitive, and collecting them might violate participants' privacy, discourage them from contributing to the study, or even introduce biases in the data collection.
(Note that while FL-based detection would ensure that raw telemetries are not disclosed, we do need them for our experiments when measuring accuracy and finding discrepancies).

As a result, we opt to collect telemetry from a selected set of websites--more specifically, the top-ranked websites according to the Chrome User Experience Report (CrUX) from April 2024~\cite{crux}.
Overall, we collect data from 3K websites sampled from the top 5K ranked websites in the CrUX report: specifically, we take the top 1K ranked websites and a random 2K sample of top 1K to 5K ranked websites.
To further limit the amount of sensitive information collected, we use Cloudflare's Domain Intelligence API~\cite{cloudflare} to detect and filter out websites containing adult or potentially harmful content.
Specifically, we exclude websites from the following categories (as defined by the Domain Intelligence API):
1) {\em Adult Themes}, 2) {\em  Gambling}, 3) {\em  Questionable Content}, 4) {\em  Security Threats}, 5) {\em Violence}, and 6) {\em Security Risks}.
Out of the top 5K ranked websites from the CrUX ranking, we excluded 948 such websites, before sampling the 3K websites that we use in our study.

\subsection{Participant Recruitment}

We recruited 30 participants (aged 18 to 60) who used the Chrome browser through the Amazon Mechanical Turk (MTurk) platform~\cite{amazonmechanicalturk}.
\final{Although participants were not restricted to a specific geographical location, prior work has found that participants on the MTurk platform are primarily based in the US~\cite{difallah2018demographics}.}
The ``Human Intelligence Task'' (HIT), i.e., the ad used to recruit users on MTurk, can be found in Appendix~\ref{appsec:hit}.

Each participant was instructed to download a Chrome extension we developed and given a password to authenticate on the extension.
As mentioned above, for privacy purposes, each participant was provided with a pre-defined set of 100 websites to visit instead of collecting data from their ``natural'' browsing sessions.
(Note that the extension only collects data from this list.)

For each website, the participants were instructed to visit at least ten sub-pages (including login pages), accept cookie banners, and solve CAPTCHAs when presented to simulate realistic browsing sessions.
The extension collected the telemetry from the websites in the background and transmitted it to our server.
Other than the website telemetry required for the purpose of fingerprinting detection, no other data (e.g., demographics, name, email) was collected. %
Upon successfully visiting all of the given websites, the participants were given a ``Task Completion Code'' by the extension, which was then used to compensate them through MTurk.
In total, the participants received an average of USD 20.33 (excluding taxes and fees), which exceeds the federal minimum wage. %

Although only a relatively small number of participants were recruited, we emphasize that the focus of this study is not to draw conclusions about the distribution or popularity of websites as visited by real users in their natural browsing behaviors.
Specifically, the intent of this study is to investigate to what extent automated crawlers miss fingerprinting scripts due to their inability to replicate real human interactions in practice.
To that end, this study could have been equivalently performed by a single participant crawling all 3K websites, but this would have made it hard to recruit a participant to complete the task successfully. %

\begin{figure}[t]
  \centering
  \includegraphics[width=\linewidth]{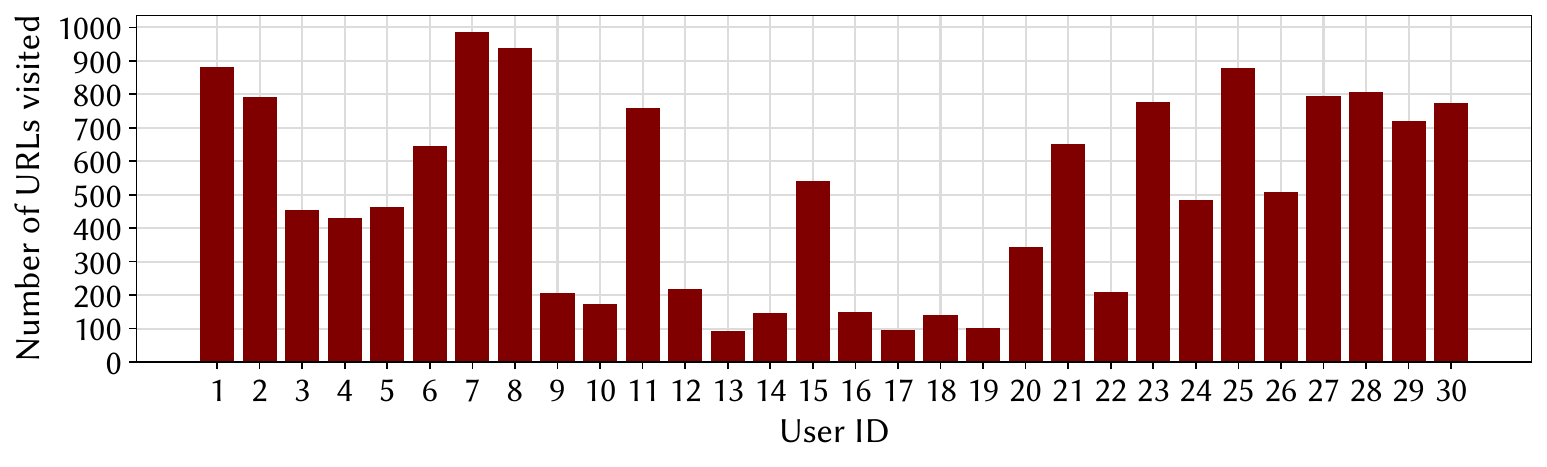}
  \caption{Distribution of unique URLs visited by each study participant.}
  \label{fig:unique_url_dist}
\end{figure}

\descr{Data Quality.}
Overall, the participants visited a total of 14,895 unique URLs.
In Figure~\ref{fig:unique_url_dist}, we plot the distribution of unique URLs visited by each participant.
While the extension verified that each participant visited all of the websites assigned to them before providing the ``Task Completion Code,'' we did not verify if each participant did visit at least ten sub-pages on each website as this would have been difficult to do so in general (e.g., single page applications may not redirect to different URLs).
On average, each participant visited 506 unique URLs.

\subsection{Collecting Scripts and Extracting Features}
Key to identifying browser fingerprinting is dynamically analyzing scripts that are loaded by the websites~\cite{ngan2022nowhere,iqbal2021fingerprinting}.
To collect this data, as mentioned, we built a Chrome extension that monitors and records the Javascript APIs accessed by each script loaded on a website, along with the associated arguments and processed return values.
Our extension then processes the collected data and sends telemetry to our server, where we then analyze it and detect fingerprinting scripts.
The extension only collects data from the domains corresponding to the list of websites assigned to each participant, i.e., it does not gather data from other websites, which the participant might visit in other browsing sessions.

We built our extension based on the instrumentation developed by Iqbal et al.~\cite{iqbal2021fingerprinting} for automated crawls with the Mozilla Firefox browser.
We adapted the instrumentation to work with the Google Chrome browser instead, extending it to monitor Chrome-specific APIs as well, and injected the instrumentation using a Chrome extension that participants can easily install.

Additionally, to reduce storage and data transmission costs, our extension first pre-processes the raw data and extracts only the features necessary to detect fingerprinting.
These features consist of API call counts (i.e., how many times each Javascript API is called) and ``custom'' features processed from the arguments and return values (e.g., length of string argument, number of elements in returned list value).
This pre-processing step not only reduces the data transmission costs but also ensures that the collected data is more private, as the exact arguments and return values are not sent to the server.
In summary, our extension processes the websites visited by the participant, the scripts loaded by each website, extracts features from each script's execution trace, and sends this data to our server for analysis.

\subsection{Fingerprinting Detection}
\label{sec:fp_detect}
We follow prior work on browser fingerprinting detection~\cite{englehardt2016online,iqbal2021fingerprinting,annamalai2024fp} and use high-precision heuristics to label scripts as {\em fingerprinting.}
Specifically, we use the heuristic developed by~\cite{englehardt2016online} and later modified by~\cite{iqbal2021fingerprinting}. %
We follow this conservative approach to labeling fingerprinting scripts over \longVer{machine learning }classifiers~\cite{iqbal2021fingerprinting} to ensure low false positives and have good confidence that the labeled scripts are, in fact, fingerprinting.
The heuristics we use identify four main types of fingerprinting -- namely,  Canvas, Canvas Font, WebRTC, and Audio Context (see Section~\ref{sec:browser_fp}  -- and does not consider simple accesses to device information as fingerprinting to minimize false positives.

\subsection{Ethics}
Our study was reviewed and approved by the UCL Computer Science Research Ethics Committee (CSREC).
\longVer{As part of the process, we submitted the full documentation detailing our approach to recruiting and compensating participants, the types of data that will be collected, along with data access policies, potential ethical issues, mitigation strategies, as well as the information sheet and consent form provided to participants.
}%
All participants were recruited and compensated anonymously through Amazon MTurk.
Specifically, each participant received an average of USD 20.33 and was conservatively expected to take at most an hour to complete visiting all 100 websites assigned to them.
Therefore, the compensation was well above minimum wages both in California~\cite{caliminimumwage} (USD 16/hour) and the United Kingdom~\cite{ukminimumwage} (GBP 11.44/hour).

\longVer{In line with data minimization principles, w}\shortVer{We} only collected pre-processed features extracted from the scripts' execution traces.
We did not collect the raw return values/arguments of the Javascript APIs or any other user data or metadata, e.g., IP address, device/network information.
As mentioned, %
we provided participants with a list of websites to visit instead of monitoring their ``natural'' browsing sessions and designed our Chrome extension to only collect data from this list.
On all other websites, the extension does not inject the instrumentation script and does not monitor or record any information from these websites.
By doing so, we can filter out potentially embarrassing or dangerous websites and prevent accidentally collecting data from participants' visits to such websites.

Additionally, the exact types of data collected along with the purpose of the data collection, potential privacy implications of taking part in the study, and participants' data rights (e.g., withdrawal from data collection) were explained in layman's terms through a participant information sheet that they could download.
Furthermore, explicit user consent was obtained through a consent form, which reiterated their rights and privacy implications of taking part.
Overall, we abided by a strict code of ethics (i.e., RESPECT~\cite{huws2004eu}).
In line with our institution's data retention policies, we will delete all data within ten years of the publication of our results. %

\section{Results}
\label{sec:results}
In this section, we present the results of our study. %
We begin by comparing the prevalence of browser fingerprinting found in real user browsing sessions with that observed in an automated crawl of the exact same \final{14.9K} websites.
Next, we shed light on why some fingerprinting websites are missed by automated crawlers, focusing on user interactions and website categories.
Finally, we compare the prevalence of Javascript APIs in fingerprinting scripts collected from real user browsing sessions with that of the automated crawl and discover potential new fingerprinting vectors.

\final{
\descr{Automated Crawl.}
Throughout this section, we compare the prevalence and distribution of fingerprinting scripts and websites present in real user browsing sessions against that of an automated crawl.
To do so, we follow the same strategy as~\citet{annamalai2024fp} and use an instrumented Chrome browser along with Puppeteer and visit the same 14,895 webpages visited by the participants during our study.
Note that this list of websites may itself not be trivial to collate as automated crawlers from prior browser fingerprinting research~\cite{annamalai2024fp,englehardt2016online,iqbal2021fingerprinting} do not visit any subpage linked from the main website.
Nevertheless, to evaluate the impact of real user interactions on browser fingerprinting beyond sub-page navigation, we assume that our automated crawler has access to the full list of websites and sub-pages crawled by real users.}

\final{
As we aim to study what type of user interactions trigger fingerprinting scripts, our automated crawler does not accept cookie consent banners.
Also note that, as bot detection techniques significantly limit the crawler's ability to visit webpages~\cite{annamalai2024fp}, we simulate some degree of user interaction (i.e., scrolling and taking a full-page screenshot) in the automated crawl and also use the \emph{puppeteer-extra-stealth-plugin}~\cite{puppeteer}.
The source code for the automated crawler can be found at \url{https://github.com/spalabucr/beyond-the-crawl}.
}

\final{
\descr{Chrome Extension Bug.}
After the data collection from real users, alas, we discovered a bug in the extension's manifest used in the real user data collection that prevented the instrumentation script from being injected into iframes, which are often used for fingerprinting.
As a result, some fingerprinting scripts and websites captured by the automated crawler were missed by the real user browsing sessions.
Nevertheless, we believe the bug only had minimal impact on our findings and in fact \emph{disadvantaged} the real user browsing sessions.
}

\subsection{Prevalence of Fingerprinting}

In total, real user browsing sessions and the automated crawl collected 80,969 and 85,853 scripts, respectively.
Due to the aforementioned bug, the automated crawl collected slightly more scripts overall.
However, out of these collected scripts, 695 were found to be fingerprinting in the real user browsing sessions, compared to only 498 in the automated crawl.

\begin{table}[t]
\small
\centering
\setlength{\tabcolsep}{7pt}
\begin{tabular}{rrrr}
\toprule
{\bf Type of FP} & {\bf Real User} & {\bf Automated} \\
\midrule
Canvas & 629 & 417 \\
Canvas Font & 40 & 41 \\
Audio & 215 & 258 \\
WebRTC & 85 & 92 \\
\midrule
{\bf Total} & 695 & 498 \\
\bottomrule
\end{tabular}
  \sspace
\caption{\final{Number of fingerprinting scripts captured by real user browsing session vs. automated crawls. {\em NB:} The same script can perform multiple types of fingerprinting.}}
\label{tab:fp_types_man_vs_aut}
\end{table}

In Table~\ref{tab:fp_types_man_vs_aut}, we report the number of each type of fingerprinting script captured by real user browsing sessions and compare it with the scripts collected from the automated crawl.
Overall, similar to the general prevalence of fingerprinting scripts, we note no major difference between the presence of each type of fingerprinting.
Specifically, in both the real user browsing sessions and the automated crawl, Canvas FP is observed to be the most popular form of fingerprinting, followed by Audio, WebRTC, and Canvas Font.
\final{Additionally, the automated crawl seemingly captured more Canvas Font, Audio, and WebRTC scripts due to the bug in the Chrome extension used to collect real user data (see Section~\ref{sec:results}).}

\final{
First, \final{1.40x} more fingerprinting scripts were detected through real-user browsing sessions than through the automated crawl.
This suggests the latter may indeed miss many fingerprinting scripts in practice, corroborating preliminary findings by~\citet{annamalai2024fp}, who show similar results from a small set of websites.
Also note that this increase is despite the decrease in the number of scripts collected from real-user browsing sessions on the same set of websites;
specifically, we observe that a slightly larger percentage of scripts are found to be fingerprinting in real user browsing sessions than the automated crawl (i.e., 0.90\% vs 0.58\%).
}

\final{
Additionally, a significant portion of fingerprinting \emph{websites} detected by the real user browsing sessions were in fact missed by the automated crawl.
Specifically, websites were flagged as `fingerprinting' if they had loaded scripts that were found to be actively fingerprinting on the website during the dynamic analysis (see Section~\ref{sec:fp_detect}).
Our analysis shows that the real user browsing sessions detected 471 such fingerprinting websites, out of which the automated crawl missed 211 (45\%).
We believe this confirms that automated crawlers cannot replicate real user interactions, which can result in the prevalence of fingerprinting being potentially underestimated in practice.
}

\subsection{Undetected Fingerprinting\longVer{ on Websites} (by Sub-Pages)}

\longVer{
\begin{table}[t]
  \centering
  \setlength{\tabcolsep}{7pt}
  \begin{tabular}{rrrrr}
    \toprule
 Failed & Auth & Content & Home &    \textbf{Total}  \\
    \midrule
 15 & 15 & 46 & 135 & {\bf 211}\\
    \bottomrule
  \end{tabular}
  \sspace
  \caption{\final{Number of fingerprinting websites undetected with automated crawl broken down by reason.}}
  \label{tab:subpages_types}
  \nspace
\end{table}}

Next, we take a closer look at the user interactions that specifically trigger fingerprinting scripts on websites.
We group the \final{211} fingerprinting websites undetected by the\longVer{ automated} crawler by the specific sub-pages that loaded fingerprinting scripts\shortVer{, and discuss below.}\longVer{. We report this in Table~\ref{tab:subpages_types} and discuss in detail below.}

\descr{Failed Visits.}
First, we find that the automated crawler failed to visit \final{15} out of the 3K websites ($\approx 1\%$).
We believe this is due to bot detectors running \textit{before} the page is loaded; in fact, the majority of these websites (\final{86.7\%}) return \texttt{4XX} errors that can be associated with bot detection scripts~\cite{annamalai2024fp}.

\descr{Authentication \& Content Pages.}
A non-negligible number of fingerprinting scripts appear on authentication pages (e.g., login, account, and sign-up pages), \final{which may require real user interactions (e.g., logging in) before fingerprinting scripts are triggered.}
As a result, the automated crawler missed 15 websites that were deploying fingerprinting scripts on the authentication page.
\final{Similarly, inner content pages may only trigger fingerprinting scripts based on user interactions (e.g., clicking on posts), it missed another 46 websites that were fingerprinting on only the content page.}

\descr{Home Pages.} The automated crawl also missed fingerprinting scripts on \final{135} home pages.
This is surprising as these scripts were not triggered even though the automated crawler successfully visited these pages.
This indicates that in some cases, even if the website is loaded successfully, specific user interactions are required to trigger fingerprinting scripts.
To investigate this more deeply, we manually visited 11 of these websites (a $\approx 10\%$ sample) to observe the specific user interactions that triggered the fingerprinting scripts.
We found that 5 out of the 11 websites deployed bot detection scripts \textit{after} the websites were loaded (as opposed to \textit{before}, which we previously categorized as a ``failed visit'').
Popular vendors for this kind of script include PerimeterX (now \url{humansecurity.com}) and \url{sift.com}.
Therefore, the fingerprinting scripts were no longer triggered because the automated crawlers were detected as bots.
Also note that four websites only start fingerprinting after user consent is received from a cookie consent banner, consistent with findings from prior work~\cite{papadogiannakis2021user}.
For the remaining two pages, we were unable to pinpoint the exact user interaction that triggered browser fingerprinting, as we could not consistently get the fingerprinting script to load.

\smallskip In summary, even if automated crawlers could crawl the Web more deeply and mimic certain user interactions, they are often caught out by advanced bot detectors and may not be able to trigger fingerprinting scripts that require specific user interactions.

\subsection{Undetected Fingerprinting\longVer{ on Websites} (by Category)}

\begin{figure}[t]
  \centering
  \includegraphics[width=\linewidth]{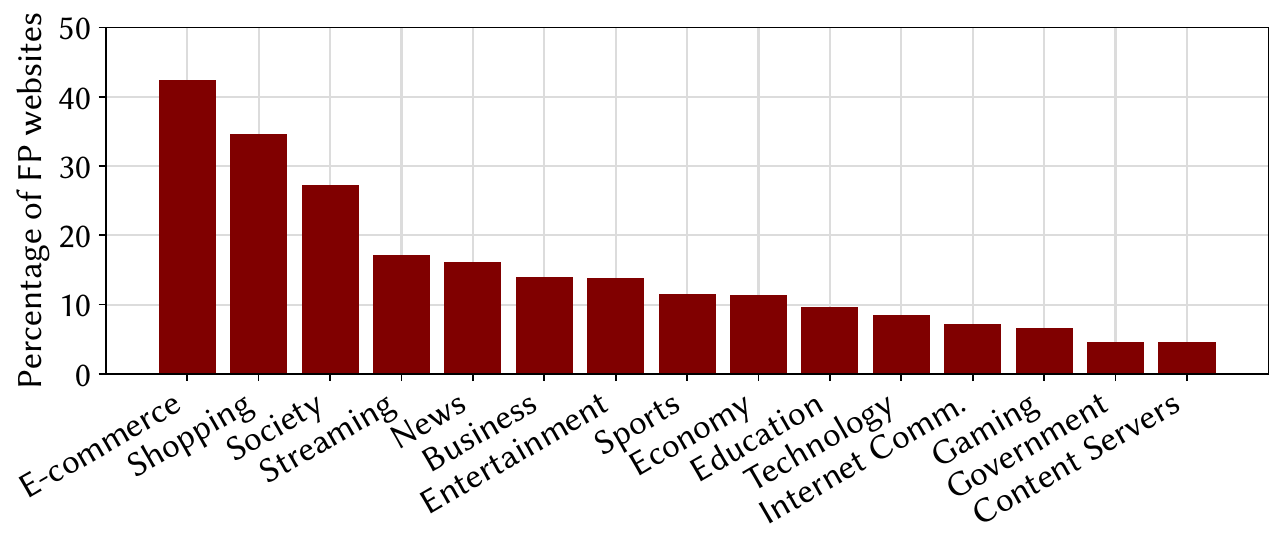}
  \caption{Percentage of fingerprinting websites for each website category.}
  \label{fig:fp_percent}
\end{figure}

Next, we analyze the prevalence of fingerprinting websites by category\footnote{We categorize the websites using the Cloudflare Domain Intelligence API~\cite{cloudflare}.} in real-user browsing sessions, as presented in Figure~\ref{fig:fp_percent}.
We also quantify the prevalence of fingerprinting by website category in the automated crawl; however, due to space limitations, we omit the results here, as we observe no significant difference between the general prevalence of fingerprinting in real-user browsing sessions and the automated crawl.

As opposed to prior work~\cite{iqbal2021fingerprinting,englehardt2016online}, we find that E-Commerce, Shopping, and Society (\& Lifestyle) categories have much higher rates of fingerprinting than News websites.
We believe one of the main reasons for this discrepancy is that previous studies used the discontinued Amazon Alexa ranking~\cite{alexa2022}, whereas we use the CrUX ranking, which is maintained and known to reflect real-user browsing patterns more accurately~\cite{ruth2022toppling}.
Additionally, as no significant difference was observed in the prevalence of fingerprinting by website category between our automated crawl and the real-user browsing sessions, we do not believe this discrepancy was due to real user behaviors.
Note that other categories (e.g., adult content) are also not represented in our results, as we had filtered out these websites for privacy reasons.

\begin{figure}[t]
  \centering
  \includegraphics[width=\linewidth]{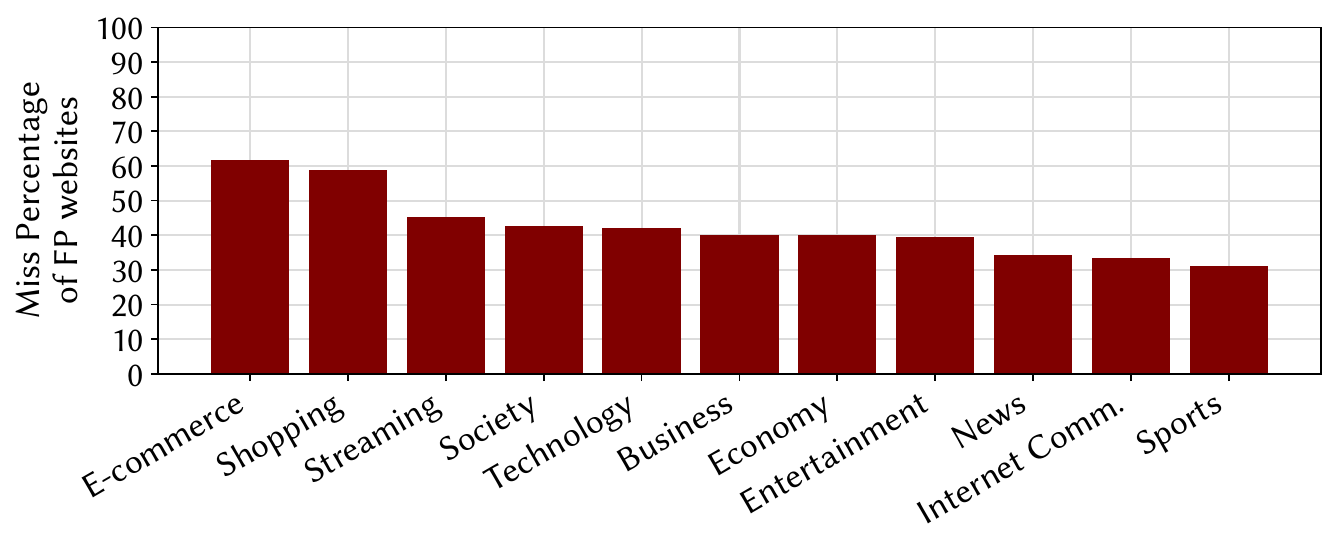}
  \caption{Percentage of fingerprinting websites undetected by the automated crawler in each website category.}
  \label{fig:miss_percent}
\end{figure}

Finally, we observe that fingerprinting is more likely to be undetected by the automated crawler for a few specific categories of websites than others.
To measure this, we introduce the notion of \textit{Miss Percentage}, defined as the percentage of websites detected as fingerprinting by the real-user browsing sessions but undetected by the automated crawl. 
Figure~\ref{fig:miss_percent} shows that \final{E-commerce, Shopping, and Video Streaming} fingerprinting websites are the most likely to be missed by automated crawlers than by real-user browsing sessions.
This is expected as \final{these types of} websites might require user interaction or user login before entering the ``main page'' of the application, where fingerprinting is expected to happen.

\subsection{Comparison of Fingerprinting APIs}

Next, we investigate the differences in the Javascript APIs used frequently by the fingerprinting scripts captured from real-user browsing sessions, as compared to those captured from automated crawls.
To do so, we quantify the relative prevalence of APIs used for fingerprinting following prior work by~\citet{iqbal2021fingerprinting} and compute the \textit{Call Ratio} for each API found in the real-user browsing sessions, i.e., the ratio between the number of times a given API is called by fingerprinting scripts and by non-fingerprinting scripts.
The Call Ratio is high (above 1) when a given API keyword is used more prevalently in fingerprinting than non-fingerprinting scripts.

Specifically, we use it to identify potential fingerprinting vectors from real-user browsing sessions that automated crawlers might otherwise miss.
In Table~\ref{tab:api_ratios}, we report the APIs with high Call Ratios in the real-user browsing sessions that were simultaneously not used by any fingerprinting scripts captured in the automated crawl (i.e., the API would have been missed by the automated crawl).
The call Ratio is $\infty$ when no non-fingerprinting scripts use the keyword.

\begin{table}[t]
\small
\centering
\setlength{\tabcolsep}{7pt}
\begin{tabular}{lr}
\toprule
{\bf Javascript API} & {\bf Call Ratio} \\
\midrule
audiocontext.sinkid                                       & $\infty$      \\
audiocontext.onsinkchange                                 & $\infty$      \\
rtcpeerconnection.getconfiguration                        & $\infty$      \\
rtcpeerconnection.sctp                                    & $\infty$      \\
rtcpeerconnection.gettransceivers                         & $\infty$      \\
rtcpeerconnection.onicecandidateerror                     & $\infty$      \\
rtcpeerconnection.tostring                                & $\infty$      \\
rtcicecandidate.address                                   & $\infty$      \\
rtcpeerconnection.addtransceiver                          & 33.0          \\
window.navigator.plugins{[}chrome pdf plugin{]}           & 7.50          \\
window.navigator.plugins{[}webkit built-in pdf{]}         & 6.77          \\
window.navigator.plugins{[}microsoft edge pdf viewer{]}   & 6.77          \\
window.navigator.plugins{[}chrome pdf viewer{]}           & 6.77          \\
window.navigator.plugins{[}chromium pdf viewer{]}         & 6.77          \\
window.navigator.plugins{[}pdf viewer{]}                  & 5.65          \\
offlineaudiocontext.hasownproperty                        & 3.62          \\
\bottomrule
\end{tabular}
  \sspace
\caption{\final{Call Ratio of a sample of Javascript APIs predominantly used by fingerprinting scripts in real-world browsing\longVer{ sessions}.}}
\label{tab:api_ratios}
\nspace
\end{table}

We observe that audio and WebRTC fingerprinting are more prevalent in real-user browsing sessions than automated crawls.
Specifically, we observe that multiple Audio APIs (i.e., audiocontext.sinkid, audiocontext.onsinkchange) and WebRTC APIs (i.e., rtcpeerconnection.getconfiguration, rtcpeerconnection.sctp, rtcpeerconnection.tostring) are exclusively used by fingerprinting scripts in real-user browsing sessions (Call Ratio = $\infty$).
At the same time, these APIs did not appear to be used by any fingerprinting script in the automated crawl.
This is probably due to these fingerprinting techniques only occurring if audio devices or peers are present in the network and triggered, which requires real-user devices with an audio interface or a crawler setup that simulates them effectively.
Nevertheless, prior work~\cite{englehardt2016online} has identified the AudioContext and RTCPeerConnection APIs as prominent fingerprinting vectors.

On the other hand, accesses to the Navigator API has previously not been considered a robust signal to detect fingerprinting~\cite{iqbal2021fingerprinting,englehardt2016online} as they are often used by non-fingerprinting scripts as well.
Conversely, our analysis of real-user browsing sessions shows that specific attributes of the Navigator API can, in fact, be used as reliable signals. 
Specifically, from Table~\ref{tab:api_ratios}, we note that the PDF viewer plugin is used predominantly by fingerprinting scripts to identify the specific browser being used.

\final{
While the Call Ratios for these APIs are relatively small compared to those in~\cite{iqbal2021fingerprinting}, this is most likely due to the small number of websites studied.
Nevertheless, they are still larger than well-known fingerprinting vectors like rtcpeerconnection.createdatachannel and audiocontext.destination, which had Call Ratios of 4.22 and 3.70, respectively.
}
Upon closer inspection, we observe that fingerprinting scripts use the ``length'' and ``description'' attributes of these plugins to identify which PDF viewer is used by the browser.
Although this API is now deprecated, it is still available in all modern browsers~\cite{navigatorplugin}; the differences in how the specification is implemented across different versions of browsers (e.g., returning a hard-coded list) might be a useful fingerprinting vector.

Overall, this confirms that real-user browsing sessions shed better light on fingerprinting vectors used in the wild compared to automated crawlers.

\begin{figure*}[t]
  \centering
  \includegraphics[width=.8\linewidth]{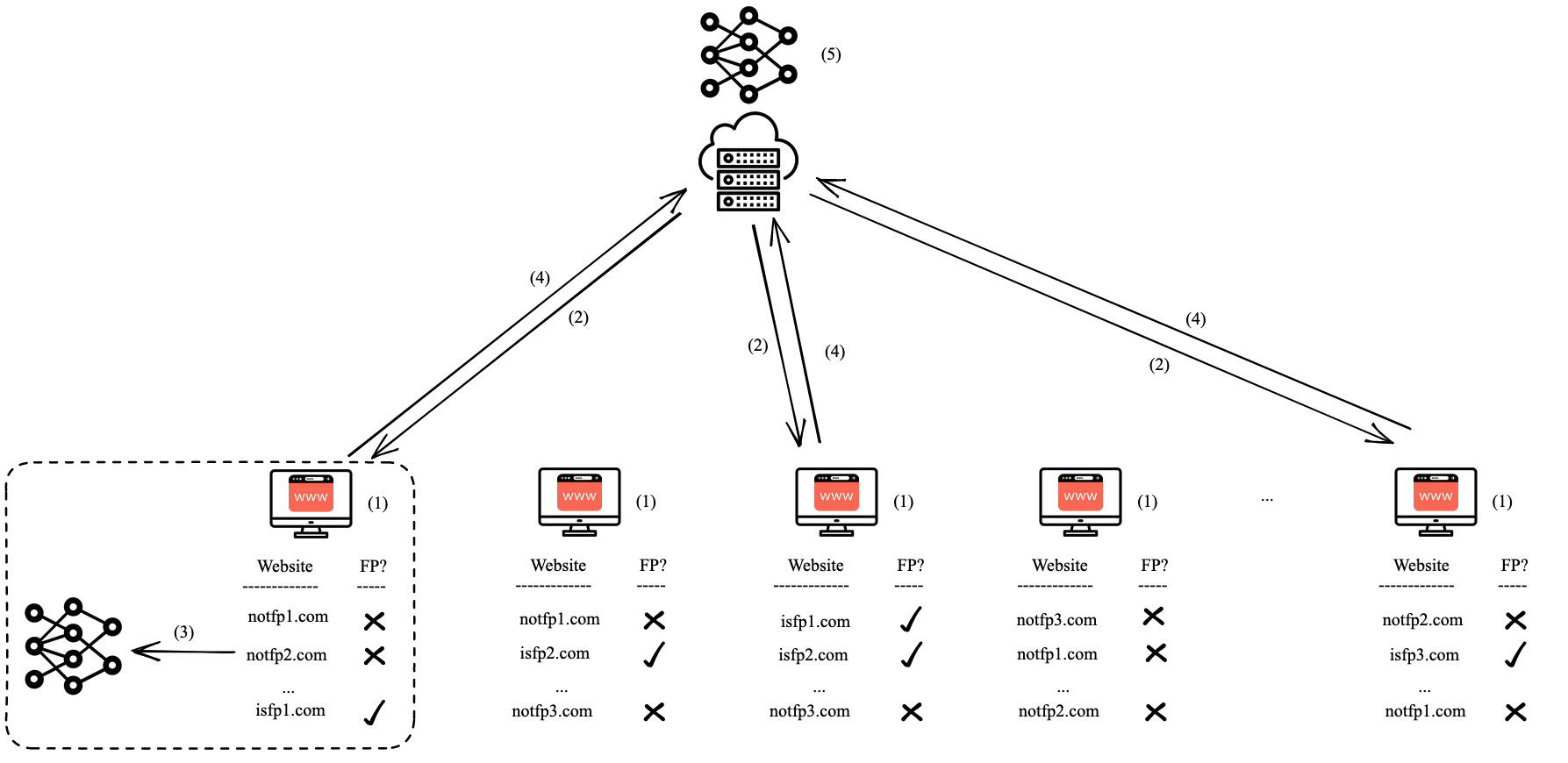}
  \caption{An overview of FP-Fed~\cite{annamalai2024fp}: (1) Participants browse websites and collect script execution data. (2) For each round of training, the server sends the previous round's global model parameters to a few selected participants. (3) These selected participants train a local model based on the collected local script execution data, and (4) send the local model updates to the server, (5) which aggregates them and adds differentially private noise to the aggregate to form the global model of the current round. Steps (2) to (5) are then repeated until the model converges.}
  \label{fig:sysdiag}
\end{figure*}

\section{Privacy-Preserving Federated Browser Fingerprinting Detection}
In the previous section, we showed that automated crawls might indeed miss fingerprinting scripts that are instead captured by real-user browsing sessions.
This sheds light on the benefits of training fingerprinting detection models on telemetry from real-user browsing sessions.
However, having users share this data with a third party would be both inefficient and extremely privacy-invasive.
On the other hand, only training locally (i.e., each user trains their own local model) would yield low accuracy and result in losing crucial knowledge about new fingerprinting behavior.

Recently,~\citet{annamalai2024fp} introduced FP-Fed, a system allowing participants to collaboratively build a browser fingerprinting detection model while keeping their data private, but only sharing model updates using the Federated Learning paradigm~\cite{mcmahan2016federated}.
However, they ultimately only tested it on an automated crawl (i.e., they did not capture real user interactions).
In the rest of this section, we investigate whether FP-Fed is effective at training browser fingerprinting detectors on real-user datasets.

\subsection{Overview of FP-Fed~\cite{annamalai2024fp}}\label{sec:fpfed}
In a nutshell, FP-Fed works as follows.
First, participants train local models to detect browser fingerprinting based on the data collected from their individual browsing sessions. 
The participants then share the model updates (but not the raw collected data) with a central server, which aggregates the model updates and adds noise to satisfy differential privacy and protect participants' privacy.
This differentially private ``global'' model is then shared with the participants, and the process repeats over multiple rounds until the global model converges.
More precisely, there are five main steps to FP-Fed, also depicted in Figure~\ref{fig:sysdiag}:
\begin{enumerate}
  \item Participants run an instrumented browser (e.g., with a Chrome extension installed) and naturally browse websites.
  \item The instrumented browser / extension collects script execution data and performs the following actions before federated training begins:
  \begin{enumerate}
    \item Collects number of times specific (often high-entropy) monitored APIs are called, along with associated return values and arguments;
    \item Extracts high-level features for each script loaded on any visited website (e.g., length of string argument, number of values in list returned, etc.);
    \item Assigns seed ground truth labels (i.e., fingerprinting/non-fingerprinting) to each script, according to a high-precision heuristic;
    \item Participates in a differentially private federated feature pre-processing phase that normalizes extracted high-level features in a privacy-preserving way.
  \end{enumerate}
  \item At each round, the server sends the previous round's global model parameters to a subset of participants selected.
  \item Participants initialize a local model with the previous round's global model parameters and train their local model with their locally collected data.
  \item Participants send the model updates to the server.
  \item The server aggregates the model updates, adds statistical noise, and computes the global model parameters for the next round.
  \item Steps 2) to 5) are repeated over multiple rounds until model convergence;
\end{enumerate}

Once the global model converges in FP-Fed, training stops, and all participants can use the trained global model for on-device browser fingerprinting detection by all participants.
FP-Fed can be run regularly (e.g., once every few weeks) so that the model can be updated to learn the latest fingerprinting behaviors. %

\descr{Improvements.}
In our work, we do not only deploy FP-Fed but also modify it to improve model performance.
Specifically, instead of training the local models from scratch, as done in~\cite{annamalai2024fp}, we first pre-train the local models, non-privately, on public data and then fine-tune them on the private local training datasets. %
This is a popular approach used to privately train ML models that yields significantly better model performance~\cite{tramer2021differentially,de2022unlocking}.

In our evaluation, we pre-train the local models on an equivalent automated crawl of the \final{14.9K} websites visited by real users.
As this is public data, pre-training can be done non-privately without degrading the privacy guarantees provided by differential privacy.
Furthermore, by combining the automated crawl and real-user browser sessions, we argue that the model can achieve precision that is close to models built only on the automated crawl from prior work~\cite{iqbal2021fingerprinting}, while learning from many more fingerprinting scripts present in the real-user browsing sessions thus improving the model's recall.
We test this claim empirically next.

\subsection{Evaluation}
\label{sec:results_fed_fp}

\begin{table*}[t]
\centering
\begin{tabular}{l|ccc|c}
\toprule
\bf Metric & \multicolumn{3}{c|}{\bf FP-Fed on Real-User Sessions} & \bf Automated \\[0.2ex]
& $\varepsilon = 1$ & $\varepsilon = 5$ & $\varepsilon = 10$ & \bf (Centralized) \\
\midrule
\bf False Pos. & 3.4 $\pm$ 2.6 & 3.0 $\pm$ 2.8 & 3.0 $\pm$ 2.8 & 7.0 $\pm$ 3.2 \\
\bf Precision & 0.98 $\pm$ 0.02 & 0.98 $\pm$ 0.02 & 0.98 $\pm$ 0.02 & 0.95 $\pm$ 0.02 \\
\bf Recall & 0.98 $\pm$ 0.01 & 0.98 $\pm$ 0.01 & 0.98 $\pm$ 0.01 & 0.95 $\pm$ 0.08 \\
\bf AUPRC & 0.98 $\pm$ 0.02 & 0.98 $\pm$ 0.02 & 0.98 $\pm$ 0.02 & 0.96 $\pm$ 0.03 \\
\bottomrule
\end{tabular}
  \sspace
\caption{\final{Performance of Fed-FP~\cite{annamalai2024fp} wrt various metrics at various levels of privacy ($\varepsilon$) on real-user browsing sessions vs.~a centralized model built on the automated crawl.}}
\label{tab:fpfed_man_vs_aut}
\nspace
\end{table*}

We now analyze the feasibility and effectiveness of training detection models for browser fingerprinting using FP-Fed~\cite{annamalai2024fp}.
To do so, we first partition the scripts from real-user browsing sessions into 80\% for training (64,565 scripts, 556 fingerprinting) and 20\% for testing (16,193 scripts, 139 fingerprinting).
Next, we re-sample the training data using the fine-grained Tranco~\cite{victor2019tranco} ranking as done in~\cite{annamalai2024fp} to simulate 1 million users participating in FP-Fed, as expected in a real-world scenario.
Finally, we train a Logistic Regression classifier using FP-Fed at various privacy levels and test the model performance on the test set.
As mentioned above, recall that we first pre-train the local models of participants with the scripts collected from the automated crawl.

In Table~\ref{tab:fpfed_man_vs_aut}, we report the Area Under Precision-Recall Curve (AUPRC) statistic along with the standard deviation over 5-fold stratified cross-validation to assess model performance.
We choose AUPRC as it summarizes the model performance over many possible thresholds that can be tuned to achieve a desired precision/recall trade-off.
\final{We also report the total number of False Positives, Precision, and Recall to provide context for the model performance at a given threshold.}
Furthermore, to assess the robustness of the model in a potential real-world deployment scenario, we compare its performance in two distinct cases. 
In the first one, we start by training a base model (centrally, no federation) with data from the automated crawl and then fine-tune it with FP-Fed on actual data from real-user browsing sessions. 
In the second case, which corresponds to the traditional approach commonly used in the literature, we train the central model with data from the automated crawl alone.
To provide a realistic performance assessment on real-world data, we evaluate both models on the 20\% testing set of the real-user browsing sessions.

Our results show that even in high privacy settings ($\varepsilon = 1.0$), the model trained on real-user browsing sessions reaches an AUPRC = \final{0.98}, with performance slightly exceeding the centralized model trained with no privacy on the automated crawl only, which achieves AUPRC = \final{0.96}.
This is notable considering that the noise introduced by DP makes it challenging for private classifiers to achieve even comparable performance to their non-private counterparts~\cite{de2022unlocking}.
In fact, even at a moderate privacy level ($\varepsilon = 5.0$), the model trained on real-user browsing sessions outperforms the automated crawl by more than one percentage point.
\final{
We observe similar trends for all the metrics, especially for the total number of false positives.
Specifically, the classifier trained on the automated crawl results in 2x as many false positives as compared to the classifiers built on the real crawl, even at high privacy regimes ($\varepsilon = 1.0$).
As mentioned in Section~\ref{sec:fp_detect}, minimizing false positives is an important aspect of building reliable browser fingerprinting detectors, for which we show that training on real user browsing sessions is crucial. 
}

\final{
One limitation of our evaluation is the large standard deviations across all metrics.
However, this is unavoidable under our evaluation setup due to the heavy imbalance in the dataset collected, which contains 80.9K scripts but only 695 FP scripts.
In the future, we aim to conduct the study at a much larger scale, which we believe will substantially mitigate this issue.
}
All in all, our experiments confirm that the federated model can leverage the fingerprinting behaviors of the scripts present in the combined browsing sessions from real users that are otherwise unavailable to models trained on the automated crawls alone.

\section{Conclusion}
\label{sec:limitations}
\longVer{This paper presented the design and execution of a user study geared to investigate the differences in the prevalence and distribution of browser fingerprinting in real-user browsing sessions as opposed to automated crawlers predominantly used by prior work in browser fingerprinting~\cite{englehardt2016online,iqbal2021fingerprinting}.
To do so, we built a Chrome extension and collected fingerprinting scripts from 30 participants as they browsed 3,000 top-ranked websites. 
We compared the resulting differences by simultaneously performing an automated crawl of the same websites.
Additionally, we evaluated the feasibility and effectiveness of collaboratively and privately training a distributed browser fingerprinting detection model using federated learning.}
\shortVer{This paper compared the prevalence and distribution of browser fingerprinting in real-user browsing sessions vs.~automated crawlers.}
Our analysis showed not only that automated crawls missed a non-negligible amount of fingerprinting scripts but that they also heavily underestimated the prevalence of browser fingerprinting in top-ranked websites.
Specifically, we observed that \final{45\%} of websites identified as fingerprinting from real-user browsing sessions were undetected by the automated crawl.
Our findings empirically validate existing hypotheses that discrepancies arise because automated crawlers lack the behavioral nuances of human users.
Consequently, bot detection scripts often block them or fail to trigger website fingerprinting mechanisms.
Finally, we showed that ML models trained on a combination of crawled data and subsequently fine-tuned in a privacy-preserving way with data from real browsing sessions on-device likely yield better performance on real-world datasets than models trained purely on crawled data.

\descr{Limitations.} \longVer{Naturally, our work is not without limitations. For instance, t}%
\final{\shortVer{T}he scope of website coverage in our study is somewhat limited, as we restrict the total number and type of websites visited by real users, which can introduce bias to the collected data -- e.g., adult entertainment websites that are well-known to fingerprint are specifically left out of data collection.
However, this was ultimately necessary not only to protect the privacy of users but also to obtain ethics approval from our Institutional Review Board.
Our data collection methodology already anonymizes users, and MTurk adds another layer of anonymity, but browsing data, especially execution traces, can still be extremely sensitive and revealing.
Therefore, this raises not only privacy but ethical concerns as well.
Nevertheless, note that our analysis already encompasses an order of magnitude more websites than previous research~\cite{annamalai2024fp}.
Moreover, we believe the study meets its primary objective, i.e., investigating discrepancies in fingerprinting prevalence and distribution between real-user browsing sessions and automated crawls.
}

\longVer{Having established a demonstrable difference, we are confident future research will expand the study's reach by increasing the number of websites analyzed, the participant pool, and the browsers covered.
Crucially, this expansion will necessitate further development of privacy-preserving data collection and analysis techniques (e.g., Differential Privacy) \final{so that we can formally guarantee the privacy of users when collecting potentially highly sensitive data at scale}, a direction we intend to explore in future work.}

\longVer{
\descr{Acknowledgements.}
This work has been supported by the National Science Scholarship (PhD) from the Agency for Science Technology and Research, Singapore.
We also wish to thank Antonis Pappasavva, Alexandros Efstratiou, and Georgi Ganev for providing their help in testing the data collection software.
}

\bibliographystyle{ACM-Reference-Format}

\begin{thebibliography}{57}


\ifx \showCODEN    \undefined \def \showCODEN     #1{\unskip}     \fi
\ifx \showDOI      \undefined \def \showDOI       #1{#1}\fi
\ifx \showISBNx    \undefined \def \showISBNx     #1{\unskip}     \fi
\ifx \showISBNxiii \undefined \def \showISBNxiii  #1{\unskip}     \fi
\ifx \showISSN     \undefined \def \showISSN      #1{\unskip}     \fi
\ifx \showLCCN     \undefined \def \showLCCN      #1{\unskip}     \fi
\ifx \shownote     \undefined \def \shownote      #1{#1}          \fi
\ifx \showarticletitle \undefined \def \showarticletitle #1{#1}   \fi
\ifx \showURL      \undefined \def \showURL       {\relax}        \fi
\providecommand\bibfield[2]{#2}
\providecommand\bibinfo[2]{#2}
\providecommand\natexlab[1]{#1}
\providecommand\showeprint[2][]{arXiv:#2}

\bibitem[Acar et~al\mbox{.}(2014)]%
        {acar2014web}
\bibfield{author}{\bibinfo{person}{Gunes Acar}, \bibinfo{person}{Christian
  Eubank}, \bibinfo{person}{Steven Englehardt}, \bibinfo{person}{Marc Juarez},
  \bibinfo{person}{Arvind Narayanan}, {and} \bibinfo{person}{Claudia Diaz}.}
  \bibinfo{year}{2014}\natexlab{}.
\newblock \showarticletitle{{The Web Never Forgets: Persistent Tracking
  Mechanisms in the Wild}}. In \bibinfo{booktitle}{\emph{{ACM CCS}}}.
\newblock


\bibitem[Acar et~al\mbox{.}(2013)]%
        {acar2013fpdetective}
\bibfield{author}{\bibinfo{person}{Gunes Acar}, \bibinfo{person}{Marc Juarez},
  \bibinfo{person}{Nick Nikiforakis}, \bibinfo{person}{Claudia Diaz},
  \bibinfo{person}{Seda G{\"u}rses}, \bibinfo{person}{Frank Piessens}, {and}
  \bibinfo{person}{Bart Preneel}.} \bibinfo{year}{2013}\natexlab{}.
\newblock \showarticletitle{{FPDetective: Dusting the Web for Fingerprinters}}.
  In \bibinfo{booktitle}{\emph{ACM CCS}}.
\newblock


\bibitem[Adnan et~al\mbox{.}(2022)]%
        {adnan2022federated}
\bibfield{author}{\bibinfo{person}{Mohammed Adnan}, \bibinfo{person}{Shivam
  Kalra}, \bibinfo{person}{Jesse~C Cresswell}, \bibinfo{person}{Graham~W
  Taylor}, {and} \bibinfo{person}{Hamid~R Tizhoosh}.}
  \bibinfo{year}{2022}\natexlab{}.
\newblock \showarticletitle{{Federated learning and differential privacy for
  medical image analysis}}.
\newblock \bibinfo{journal}{\emph{Scientific reports}} \bibinfo{volume}{12},
  \bibinfo{number}{1} (\bibinfo{year}{2022}).
\newblock


\bibitem[Akhavani et~al\mbox{.}(2021)]%
        {akhavani2021browserprint}
\bibfield{author}{\bibinfo{person}{Seyed~Ali Akhavani}, \bibinfo{person}{Jordan
  Jueckstock}, \bibinfo{person}{Junhua Su}, \bibinfo{person}{Alexandros
  Kapravelos}, \bibinfo{person}{Engin Kirda}, {and} \bibinfo{person}{Long Lu}.}
  \bibinfo{year}{2021}\natexlab{}.
\newblock \showarticletitle{{Browserprint: An Analysis of the Impact of Browser
  Features on Fingerprintability and Web Privacy}}. In
  \bibinfo{booktitle}{\emph{Information Security Conference}}.
\newblock


\bibitem[Alaca and Van~Oorschot(2016)]%
        {alaca2016device}
\bibfield{author}{\bibinfo{person}{Furkan Alaca} {and} \bibinfo{person}{Paul~C
  Van~Oorschot}.} \bibinfo{year}{2016}\natexlab{}.
\newblock \showarticletitle{{Device fingerprinting for augmenting web
  authentication: classification and analysis of methods}}. In
  \bibinfo{booktitle}{\emph{{ACM CCS}}}.
\newblock


\bibitem[Amazon(2005)]%
        {amazonmechanicalturk}
\bibfield{author}{\bibinfo{person}{Amazon}.} \bibinfo{year}{2005}\natexlab{}.
\newblock \bibinfo{title}{{Amazon Mechanical Turk}}.
\newblock \bibinfo{howpublished}{\url{https://www.mturk.com/}}.
\newblock


\bibitem[Amazon(2022)]%
        {alexa2022}
\bibfield{author}{\bibinfo{person}{Amazon}.} \bibinfo{year}{2022}\natexlab{}.
\newblock \bibinfo{title}{{We will be retiring Alexa.com on May 1, 2022}}.
\newblock
  \bibinfo{howpublished}{\url{https://web.archive.org/web/20220102200605/https://support.alexa.com/hc/en-us/articles/4410503838999}}.
\newblock


\bibitem[Annamalai et~al\mbox{.}(2024)]%
        {annamalai2024fp}
\bibfield{author}{\bibinfo{person}{Meenatchi Sundaram Muthu~Selva Annamalai},
  \bibinfo{person}{Igor Bilogrevic}, {and} \bibinfo{person}{Emiliano
  De~Cristofaro}.} \bibinfo{year}{2024}\natexlab{}.
\newblock \showarticletitle{{FP-Fed: Privacy-Preserving Federated Detection of
  Browser Fingerprinting}}.
\newblock \bibinfo{journal}{\emph{NDSS}}.
\newblock


\bibitem[Apple(2023)]%
        {applefp}
\bibfield{author}{\bibinfo{person}{Apple}.} \bibinfo{year}{2023}\natexlab{}.
\newblock \bibinfo{title}{{Apple announces powerful new privacy and security
  features}}.
\newblock
  \bibinfo{howpublished}{\url{https://www.apple.com/sg/newsroom/2023/06/apple-announces-powerful-new-privacy-and-security-features/}}.
\newblock


\bibitem[Bahrami et~al\mbox{.}(2022)]%
        {bahrami2021}
\bibfield{author}{\bibinfo{person}{Pouneh~Nikkhah Bahrami},
  \bibinfo{person}{Umar Iqbal}, {and} \bibinfo{person}{Zubair Shafiq}.}
  \bibinfo{year}{2022}\natexlab{}.
\newblock \showarticletitle{{FP-Radar: Longitudinal Measurement and Early
  Detection of Browser Fingerprinting}}.
\newblock \bibinfo{journal}{\emph{PETS}} (\bibinfo{year}{2022}).
\newblock


\bibitem[Berstend(2023)]%
        {puppeteer}
\bibfield{author}{\bibinfo{person}{Berstend}.} \bibinfo{year}{2023}\natexlab{}.
\newblock \bibinfo{title}{{puppeteer-extra-plugin-stealth}}.
\newblock
  \bibinfo{howpublished}{\url{https://github.com/berstend/puppeteer-extra}}.
\newblock


\bibitem[Bojinov et~al\mbox{.}(2014)]%
        {bojinov2014mobile}
\bibfield{author}{\bibinfo{person}{Hristo Bojinov}, \bibinfo{person}{Yan
  Michalevsky}, \bibinfo{person}{Gabi Nakibly}, {and} \bibinfo{person}{Dan
  Boneh}.} \bibinfo{year}{2014}\natexlab{}.
\newblock \showarticletitle{{Mobile Device Identification via Sensor
  Fingerprinting}}.
\newblock \bibinfo{journal}{\emph{arXiv:1408.1416}} (\bibinfo{year}{2014}).
\newblock


\bibitem[Brave(2024)]%
        {bravefp}
\bibfield{author}{\bibinfo{person}{Brave}.} \bibinfo{year}{2024}\natexlab{}.
\newblock \bibinfo{title}{{Fingerprinting Protections}}.
\newblock
  \bibinfo{howpublished}{\url{https://github.com/brave/brave-browser/wiki/Fingerprinting-Protections}}.
\newblock


\bibitem[Cao et~al\mbox{.}(2017)]%
        {cao2017cross}
\bibfield{author}{\bibinfo{person}{Yinzhi Cao}, \bibinfo{person}{Song Li},
  {and} \bibinfo{person}{Erik Wijmans}.} \bibinfo{year}{2017}\natexlab{}.
\newblock \showarticletitle{{(Cross-)Browser Fingerprinting via OS and Hardware
  Level Features}}. In \bibinfo{booktitle}{\emph{NDSS}}.
\newblock


\bibitem[Chrome(2024)]%
        {crux}
\bibfield{author}{\bibinfo{person}{Chrome}.} \bibinfo{year}{2024}\natexlab{}.
\newblock \bibinfo{title}{{Chrome User Experience Report}}.
\newblock
  \bibinfo{howpublished}{\url{https://developer.chrome.com/docs/crux/}}.
\newblock


\bibitem[Cloudflare(2024)]%
        {cloudflare}
\bibfield{author}{\bibinfo{person}{Cloudflare}.}
  \bibinfo{year}{2024}\natexlab{}.
\newblock \bibinfo{title}{{Get Domain Details}}.
\newblock
  \bibinfo{howpublished}{\url{https://developers.cloudflare.com/api/operations/domain-intelligence-get-domain-details}}.
\newblock


\bibitem[Commission(2024)]%
        {ukminimumwage}
\bibfield{author}{\bibinfo{person}{Low~Pay Commission}.}
  \bibinfo{year}{2024}\natexlab{}.
\newblock \bibinfo{title}{{National Minimum Wage in 2024}}.
\newblock
  \bibinfo{howpublished}{\url{https://www.gov.uk/government/publications/the-national-minimum-wage-in-2024}}.
\newblock


\bibitem[Crouch(2018)]%
        {mozillabreakage}
\bibfield{author}{\bibinfo{person}{Luke Crouch}.}
  \bibinfo{year}{2018}\natexlab{}.
\newblock \bibinfo{title}{{Improving privacy without breaking the web}}.
\newblock
  \bibinfo{howpublished}{\url{https://blog.mozilla.org/data/2018/01/26/improving-privacy-without-breaking-the-web/}}.
\newblock


\bibitem[Das et~al\mbox{.}(2018)]%
        {das2018web}
\bibfield{author}{\bibinfo{person}{Anupam Das}, \bibinfo{person}{Gunes Acar},
  \bibinfo{person}{Nikita Borisov}, {and} \bibinfo{person}{Amogh Pradeep}.}
  \bibinfo{year}{2018}\natexlab{}.
\newblock \showarticletitle{{The Web's Sixth Sense: A Study of Scripts
  Accessing Smartphone Sensors}}. In \bibinfo{booktitle}{\emph{{ACM CCS}}}.
\newblock


\bibitem[De et~al\mbox{.}(2022)]%
        {de2022unlocking}
\bibfield{author}{\bibinfo{person}{Soham De}, \bibinfo{person}{Leonard
  Berrada}, \bibinfo{person}{Jamie Hayes}, \bibinfo{person}{Samuel~L Smith},
  {and} \bibinfo{person}{Borja Balle}.} \bibinfo{year}{2022}\natexlab{}.
\newblock \showarticletitle{{Unlocking High-Accuracy Differentially Private
  Image Classification through Scale}}.
\newblock \bibinfo{journal}{\emph{arXiv:2204.13650}} (\bibinfo{year}{2022}).
\newblock


\bibitem[Difallah et~al\mbox{.}(2018)]%
        {difallah2018demographics}
\bibfield{author}{\bibinfo{person}{Djellel Difallah}, \bibinfo{person}{Elena
  Filatova}, {and} \bibinfo{person}{Panos Ipeirotis}.}
  \bibinfo{year}{2018}\natexlab{}.
\newblock \showarticletitle{{Demographics and Dynamics of Mechanical Turk
  Workers}}. In \bibinfo{booktitle}{\emph{ACM International Conference on Web
  Search and Data Mining}}.
\newblock


\bibitem[Disconnect(2018)]%
        {disconnect2018}
\bibfield{author}{\bibinfo{person}{Disconnect}.}
  \bibinfo{year}{2018}\natexlab{}.
\newblock \bibinfo{title}{{Disconnect defends the digital you}}.
\newblock \bibinfo{howpublished}{\url{https://disconnect.me}}.
\newblock


\bibitem[Docs(2024)]%
        {navigatorplugin}
\bibfield{author}{\bibinfo{person}{MDN~Web Docs}.}
  \bibinfo{year}{2024}\natexlab{}.
\newblock \bibinfo{title}{{Navigator: plugins property}}.
\newblock
  \bibinfo{howpublished}{\url{https://developer.mozilla.org/en-US/docs/Web/API/Navigator/plugins}}.
\newblock


\bibitem[Dwork and Roth(2014)]%
        {dwork2014algorithmic}
\bibfield{author}{\bibinfo{person}{Cynthia Dwork} {and} \bibinfo{person}{Aaron
  Roth}.} \bibinfo{year}{2014}\natexlab{}.
\newblock \showarticletitle{{The Algorithmic Foundations of Differential
  Privacy}}.
\newblock \bibinfo{journal}{\emph{Foundations and Trends in Theoretical
  Computer Science}} (\bibinfo{year}{2014}).
\newblock


\bibitem[EasyList(2024)]%
        {easyprivacy}
\bibfield{author}{\bibinfo{person}{EasyList}.} \bibinfo{year}{2024}\natexlab{}.
\newblock \bibinfo{title}{{EasyPrivacy}}.
\newblock
  \bibinfo{howpublished}{\url{https://easylist.to/easylist/easyprivacy.txt}}.
\newblock


\bibitem[Eckersley(2010)]%
        {eckersley2010unique}
\bibfield{author}{\bibinfo{person}{Peter Eckersley}.}
  \bibinfo{year}{2010}\natexlab{}.
\newblock \showarticletitle{{How Unique Is Your Web Browser?}}. In
  \bibinfo{booktitle}{\emph{PETS}}.
\newblock


\bibitem[{EFF}(2023)]%
        {privacybadger}
\bibfield{author}{\bibinfo{person}{{EFF}}.} \bibinfo{year}{2023}\natexlab{}.
\newblock \bibinfo{title}{{{PrivacyBadger}}}.
\newblock
  \bibinfo{howpublished}{\url{https://github.com/EFForg/privacybadgerfirefox/blob/master/data/cookieblocklist.txt}}.
\newblock


\bibitem[Englehardt(2020)]%
        {firefoxbrowserfp2020}
\bibfield{author}{\bibinfo{person}{Steven Englehardt}.}
  \bibinfo{year}{2020}\natexlab{}.
\newblock \bibinfo{title}{{Firefox 72 blocks third-party fingerprinting
  resources}}.
\newblock
  \bibinfo{howpublished}{\url{https://blog.mozilla.org/security/2020/01/07/firefox-72-fingerprinting/}}.
\newblock


\bibitem[Englehardt and Narayanan(2016)]%
        {englehardt2016online}
\bibfield{author}{\bibinfo{person}{Steven Englehardt} {and}
  \bibinfo{person}{Arvind Narayanan}.} \bibinfo{year}{2016}\natexlab{}.
\newblock \showarticletitle{{Online tracking: A 1-million-site Measurement and
  Analysis}}. In \bibinfo{booktitle}{\emph{{ACM CCS}}}.
\newblock


\bibitem[Huws et~al\mbox{.}(2004)]%
        {huws2004eu}
\bibfield{author}{\bibinfo{person}{U Huws}, \bibinfo{person}{S Dench}, {and}
  \bibinfo{person}{Ron Iphofen}.} \bibinfo{year}{2004}\natexlab{}.
\newblock \bibinfo{booktitle}{\emph{{An EU code of ethics for socio-economic
  research}}}.
\newblock \bibinfo{publisher}{Institute for Employment Studies}.
\newblock


\bibitem[Ikram et~al\mbox{.}(2017)]%
        {ikram2017towards}
\bibfield{author}{\bibinfo{person}{Muhammad Ikram},
  \bibinfo{person}{Hassan~Jameel Asghar}, \bibinfo{person}{Mohamed~Ali Kaafar},
  \bibinfo{person}{Balachander Krishnamurthy}, {and} \bibinfo{person}{Anirban
  Mahanti}.} \bibinfo{year}{2017}\natexlab{}.
\newblock \showarticletitle{{Towards Seamless Tracking-Free Web: Improved
  Detection of Trackers via One-class Learning}}.
\newblock \bibinfo{journal}{\emph{PETS}} (\bibinfo{year}{2017}).
\newblock


\bibitem[Iovation(2019)]%
        {iovationfraud2019}
\bibfield{author}{\bibinfo{person}{Iovation}.} \bibinfo{year}{2019}\natexlab{}.
\newblock \bibinfo{title}{{Iovation Fraud Protection}}.
\newblock
  \bibinfo{howpublished}{\url{https://web.archive.org/web/20191130164107/\\https://www.iovation.com/fraudforce-fraud-detection-prevention}}.
\newblock


\bibitem[Iqbal et~al\mbox{.}(2021)]%
        {iqbal2021fingerprinting}
\bibfield{author}{\bibinfo{person}{Umar Iqbal}, \bibinfo{person}{Steven
  Englehardt}, {and} \bibinfo{person}{Zubair Shafiq}.}
  \bibinfo{year}{2021}\natexlab{}.
\newblock \showarticletitle{{Fingerprinting the fingerprinters: Learning to
  detect browser fingerprinting behaviors}}. In \bibinfo{booktitle}{\emph{{IEEE
  S\&P}}}.
\newblock


\bibitem[Kairouz et~al\mbox{.}(2021)]%
        {kairouz2021distributed}
\bibfield{author}{\bibinfo{person}{Peter Kairouz}, \bibinfo{person}{Ziyu Liu},
  {and} \bibinfo{person}{Thomas Steinke}.} \bibinfo{year}{2021}\natexlab{}.
\newblock \showarticletitle{{The Distributed Discrete Gaussian Mechanism for
  Federated Learning with Secure Aggregation}}. In
  \bibinfo{booktitle}{\emph{ICML}}.
\newblock


\bibitem[Laperdrix et~al\mbox{.}(2019)]%
        {laperdrix2019morellian}
\bibfield{author}{\bibinfo{person}{Pierre Laperdrix}, \bibinfo{person}{Gildas
  Avoine}, \bibinfo{person}{Benoit Baudry}, {and} \bibinfo{person}{Nick
  Nikiforakis}.} \bibinfo{year}{2019}\natexlab{}.
\newblock \showarticletitle{{Morellian analysis for browsers: Making web
  authentication stronger with canvas fingerprinting}}. In
  \bibinfo{booktitle}{\emph{{Detection of Intrusions and Malware, and
  Vulnerability Assessment}}}.
\newblock


\bibitem[Liu et~al\mbox{.}(2024)]%
        {liu2024identified}
\bibfield{author}{\bibinfo{person}{Zengrui Liu}, \bibinfo{person}{Jimmy Dani},
  \bibinfo{person}{Shujiang Wu}, \bibinfo{person}{Yinzhi Cao}, {and}
  \bibinfo{person}{Nitesh Saxena}.} \bibinfo{year}{2024}\natexlab{}.
\newblock \showarticletitle{{Identified-and-Targeted: The First Early Evidence
  of the Privacy-Invasive Use of Browser Fingerprinting for Online Tracking}}.
\newblock \bibinfo{journal}{\emph{arXiv:2409.15656}} (\bibinfo{year}{2024}).
\newblock


\bibitem[Lunden(2018)]%
        {relix2018}
\bibfield{author}{\bibinfo{person}{Ingrid Lunden}.}
  \bibinfo{year}{2018}\natexlab{}.
\newblock \bibinfo{title}{{Relx acquires ThreatMetrix for \$817M to ramp up in
  risk-based authentication}}.
\newblock
  \bibinfo{howpublished}{\url{https://techcrunch.com/2018/01/29/relx-threatmetrix-risk-authentication-lexisnexis/?guccounter=1}}.
\newblock


\bibitem[Mayer(2009)]%
        {mayer2009any}
\bibfield{author}{\bibinfo{person}{Jonathan~R Mayer}.}
  \bibinfo{year}{2009}\natexlab{}.
\newblock \showarticletitle{{``Any person... a pamphleteer'': Internet
  Anonymity in the Age of Web 2.0}}.
\newblock \bibinfo{journal}{\emph{Undergraduate Senior Thesis, Princeton
  University}}  \bibinfo{volume}{85} (\bibinfo{year}{2009}).
\newblock


\bibitem[McMahan et~al\mbox{.}(2016)]%
        {mcmahan2016federated}
\bibfield{author}{\bibinfo{person}{H~Brendan McMahan}, \bibinfo{person}{Eider
  Moore}, \bibinfo{person}{Daniel Ramage}, {and}
  \bibinfo{person}{Blaise~Ag{\"u}era y Arcas}.}
  \bibinfo{year}{2016}\natexlab{}.
\newblock \showarticletitle{{Federated learning of deep networks using model
  averaging}}.
\newblock \bibinfo{journal}{\emph{arXiv:1602.05629}} (\bibinfo{year}{2016}).
\newblock


\bibitem[McMahan et~al\mbox{.}(2018)]%
        {mcmahan2018learning}
\bibfield{author}{\bibinfo{person}{H~Brendan McMahan}, \bibinfo{person}{Daniel
  Ramage}, \bibinfo{person}{Kunal Talwar}, {and} \bibinfo{person}{Li Zhang}.}
  \bibinfo{year}{2018}\natexlab{}.
\newblock \showarticletitle{{Learning Differentially Private Recurrent Language
  Models}}. In \bibinfo{booktitle}{\emph{ICLR}}.
\newblock


\bibitem[Melis et~al\mbox{.}(2019)]%
        {melis2019exploiting}
\bibfield{author}{\bibinfo{person}{Luca Melis}, \bibinfo{person}{Congzheng
  Song}, \bibinfo{person}{Emiliano De~Cristofaro}, {and}
  \bibinfo{person}{Vitaly Shmatikov}.} \bibinfo{year}{2019}\natexlab{}.
\newblock \showarticletitle{{Exploiting Unintended Feature Leakage in
  Collaborative Learning}}. In \bibinfo{booktitle}{\emph{IEEE S\&P}}.
\newblock


\bibitem[Mowery and Shacham(2012)]%
        {mowery2012pixel}
\bibfield{author}{\bibinfo{person}{Keaton Mowery} {and} \bibinfo{person}{Hovav
  Shacham}.} \bibinfo{year}{2012}\natexlab{}.
\newblock \showarticletitle{{Pixel perfect: Fingerprinting canvas in HTML5}}.
\newblock \bibinfo{journal}{\emph{W2SP}} (\bibinfo{year}{2012}).
\newblock


\bibitem[Mozilla(2023)]%
        {webrtc}
\bibfield{author}{\bibinfo{person}{Mozilla}.} \bibinfo{year}{2023}\natexlab{}.
\newblock \bibinfo{title}{{WebRTC API}}.
\newblock
  \bibinfo{howpublished}{\url{https://developer.mozilla.org/en-US/docs/Web/API/WebRTC_API}}.
\newblock


\bibitem[Naseri et~al\mbox{.}(2022)]%
        {naseri2022cerberus}
\bibfield{author}{\bibinfo{person}{Mohammad Naseri}, \bibinfo{person}{Yufei
  Han}, \bibinfo{person}{Enrico Mariconti}, \bibinfo{person}{Yun Shen},
  \bibinfo{person}{Gianluca Stringhini}, {and} \bibinfo{person}{Emiliano
  De~Cristofaro}.} \bibinfo{year}{2022}\natexlab{}.
\newblock \showarticletitle{{Cerberus: Exploring Federated Prediction of
  Security Events}}. In \bibinfo{booktitle}{\emph{ACM CCS}}.
\newblock


\bibitem[Ngan et~al\mbox{.}(2022)]%
        {ngan2022nowhere}
\bibfield{author}{\bibinfo{person}{Ray Ngan}, \bibinfo{person}{Surya
  Konkimalla}, {and} \bibinfo{person}{Zubair Shafiq}.}
  \bibinfo{year}{2022}\natexlab{}.
\newblock \showarticletitle{{Nowhere to Hide: Detecting Obfuscated
  Fingerprinting Scripts}}.
\newblock \bibinfo{journal}{\emph{arXiv:2206.13599}} (\bibinfo{year}{2022}).
\newblock


\bibitem[of~Labor(2024)]%
        {caliminimumwage}
\bibfield{author}{\bibinfo{person}{Department of Labor}.}
  \bibinfo{year}{2024}\natexlab{}.
\newblock \bibinfo{title}{{State Minimum Wage Laws}}.
\newblock
  \bibinfo{howpublished}{\url{https://www.dol.gov/agencies/whd/minimum-wage/state}}.
\newblock


\bibitem[Olejnik et~al\mbox{.}(2016)]%
        {olejnik2016leaking}
\bibfield{author}{\bibinfo{person}{{\L}ukasz Olejnik}, \bibinfo{person}{Gunes
  Acar}, \bibinfo{person}{Claude Castelluccia}, {and} \bibinfo{person}{Claudia
  Diaz}.} \bibinfo{year}{2016}\natexlab{}.
\newblock \showarticletitle{{The leaking battery: A privacy analysis of the
  HTML5 Battery Status API}}. In \bibinfo{booktitle}{\emph{DPM}}.
\newblock


\bibitem[Papadogiannakis et~al\mbox{.}(2021)]%
        {papadogiannakis2021user}
\bibfield{author}{\bibinfo{person}{Emmanouil Papadogiannakis},
  \bibinfo{person}{Panagiotis Papadopoulos}, \bibinfo{person}{Nicolas
  Kourtellis}, {and} \bibinfo{person}{Evangelos~P Markatos}.}
  \bibinfo{year}{2021}\natexlab{}.
\newblock \showarticletitle{{User Tracking in the Post-cookie Era: How Websites
  Bypass GDPR Consent to Track Users}}. In \bibinfo{booktitle}{\emph{WWW}}.
\newblock


\bibitem[Pochat et~al\mbox{.}(2019)]%
        {victor2019tranco}
\bibfield{author}{\bibinfo{person}{Victor~Le Pochat}, \bibinfo{person}{Tom van
  Goethem}, \bibinfo{person}{Samaneh Tajalizadehkhoob}, \bibinfo{person}{Maciej
  Korczynski}, {and} \bibinfo{person}{Wouter Joosen}.}
  \bibinfo{year}{2019}\natexlab{}.
\newblock \showarticletitle{{Tranco: A Research-Oriented Top Sites Ranking
  Hardened Against Manipulation}}. In \bibinfo{booktitle}{\emph{NDSS}}.
\newblock


\bibitem[Pugliese et~al\mbox{.}(2020)]%
        {pugliese2020long}
\bibfield{author}{\bibinfo{person}{Gaston Pugliese}, \bibinfo{person}{Christian
  Riess}, \bibinfo{person}{Freya Gassmann}, {and} \bibinfo{person}{Zinaida
  Benenson}.} \bibinfo{year}{2020}\natexlab{}.
\newblock \showarticletitle{{Long-Term Observation on Browser Fingerprinting:
  Users' Trackability and Perspective}}.
\newblock \bibinfo{journal}{\emph{PETS}} (\bibinfo{year}{2020}).
\newblock


\bibitem[Roesner et~al\mbox{.}(2012)]%
        {roesner2012detecting}
\bibfield{author}{\bibinfo{person}{Franziska Roesner},
  \bibinfo{person}{Tadayoshi Kohno}, {and} \bibinfo{person}{David Wetherall}.}
  \bibinfo{year}{2012}\natexlab{}.
\newblock \showarticletitle{{Detecting and Defending Against Third-Party
  Tracking on the Web}}. In \bibinfo{booktitle}{\emph{USENIX}}.
\newblock


\bibitem[Ruth et~al\mbox{.}(2022)]%
        {ruth2022toppling}
\bibfield{author}{\bibinfo{person}{Kimberly Ruth}, \bibinfo{person}{Deepak
  Kumar}, \bibinfo{person}{Brandon Wang}, \bibinfo{person}{Luke Valenta}, {and}
  \bibinfo{person}{Zakir Durumeric}.} \bibinfo{year}{2022}\natexlab{}.
\newblock \showarticletitle{{Toppling Top Lists: Evaluating the Accuracy of
  Popular Website Lists}}. In \bibinfo{booktitle}{\emph{ACM IMC}}.
\newblock


\bibitem[Senol et~al\mbox{.}(2024)]%
        {senol2024double}
\bibfield{author}{\bibinfo{person}{Asuman Senol}, \bibinfo{person}{Alisha
  Ukani}, \bibinfo{person}{Dylan Cutler}, {and} \bibinfo{person}{Igor
  Bilogrevic}.} \bibinfo{year}{2024}\natexlab{}.
\newblock \showarticletitle{{The Double Edged Sword: Identifying Authentication
  Pages and their Fingerprinting Behavior}}. In \bibinfo{booktitle}{\emph{ACM
  WWW}}.
\newblock


\bibitem[Sun et~al\mbox{.}(2021)]%
        {sun2021ldp}
\bibfield{author}{\bibinfo{person}{Lichao Sun}, \bibinfo{person}{Jianwei Qian},
  {and} \bibinfo{person}{Xun Chen}.} \bibinfo{year}{2021}\natexlab{}.
\newblock \showarticletitle{{LDP-FL: Practical Private Aggregation in Federated
  Learning with Local Differential Privacy}}. In
  \bibinfo{booktitle}{\emph{IJCAI}}.
\newblock


\bibitem[Tramer and Boneh(2021)]%
        {tramer2021differentially}
\bibfield{author}{\bibinfo{person}{Florian Tramer} {and} \bibinfo{person}{Dan
  Boneh}.} \bibinfo{year}{2021}\natexlab{}.
\newblock \showarticletitle{{Differentially Private Learning Needs Better
  Features (or Much More Data)}}. In \bibinfo{booktitle}{\emph{ICLR}}.
\newblock


\bibitem[Truex et~al\mbox{.}(2020)]%
        {truex2020ldp}
\bibfield{author}{\bibinfo{person}{Stacey Truex}, \bibinfo{person}{Ling Liu},
  \bibinfo{person}{Ka-Ho Chow}, \bibinfo{person}{Mehmet~Emre Gursoy}, {and}
  \bibinfo{person}{Wenqi Wei}.} \bibinfo{year}{2020}\natexlab{}.
\newblock \showarticletitle{{LDP-Fed: Federated learning with local
  differential privacy}}. In \bibinfo{booktitle}{\emph{ACM International
  Workshop on Edge Systems, Analytics and Networking}}.
\newblock


\bibitem[W3C(2021)]%
        {w3cbrowserfp}
\bibfield{author}{\bibinfo{person}{W3C}.} \bibinfo{year}{2021}\natexlab{}.
\newblock \bibinfo{title}{{Mitigating Browser Fingerprinting in Web
  Specifications}}.
\newblock
  \bibinfo{howpublished}{\url{https://w3c.github.io/fingerprinting-guidance/}}.
\newblock


\end{thebibliography}

%
\appendix

\section{Human Intelligence Task (HIT)}
\label{appsec:hit}

\begin{figure*}[t]
  \centering
  \begin{subfigure}{0.5\textwidth}
      \includegraphics[width=\textwidth]{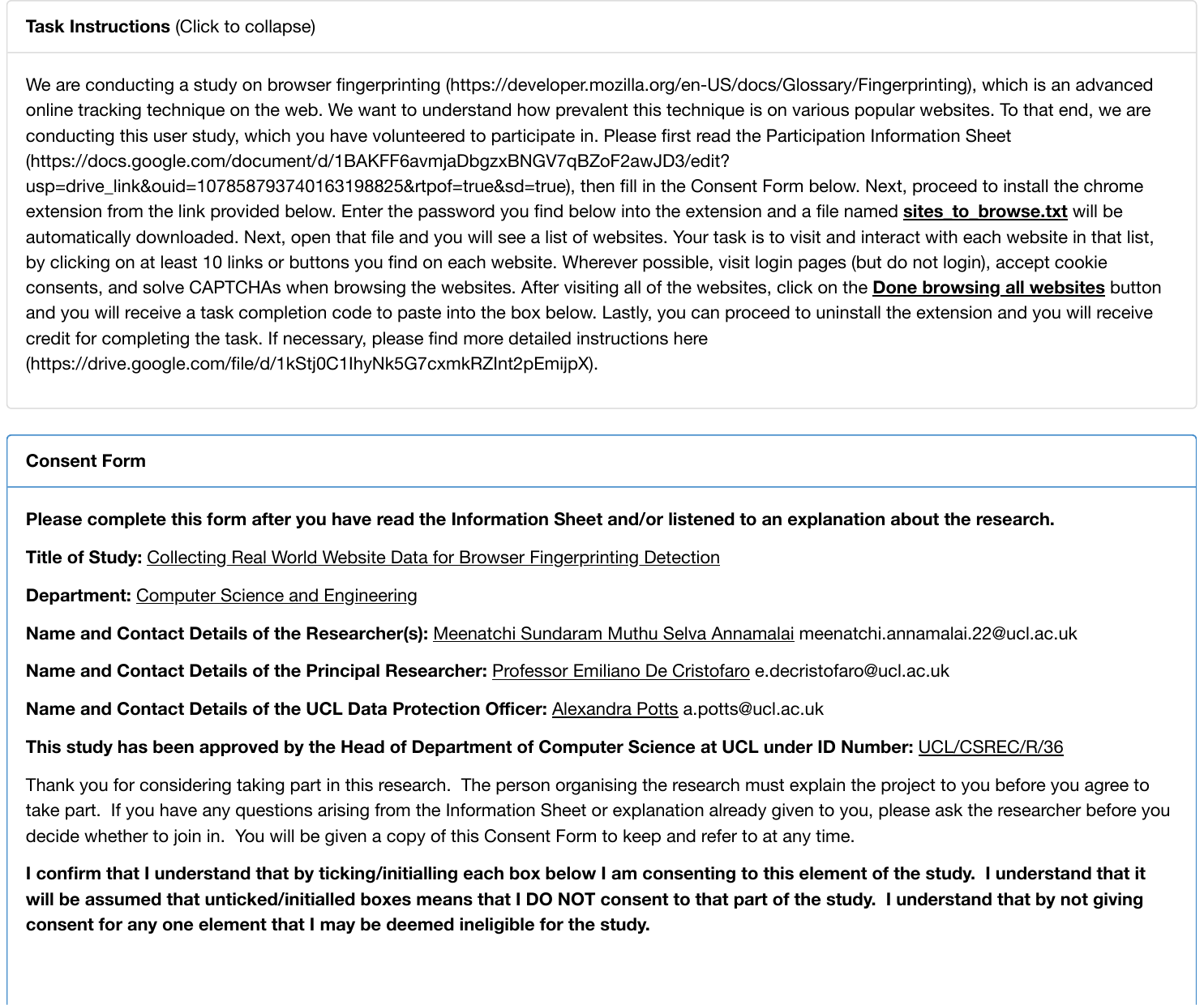}
  \end{subfigure}\\[0.5ex]
  \begin{subfigure}{0.5\textwidth}
      \includegraphics[width=\textwidth]{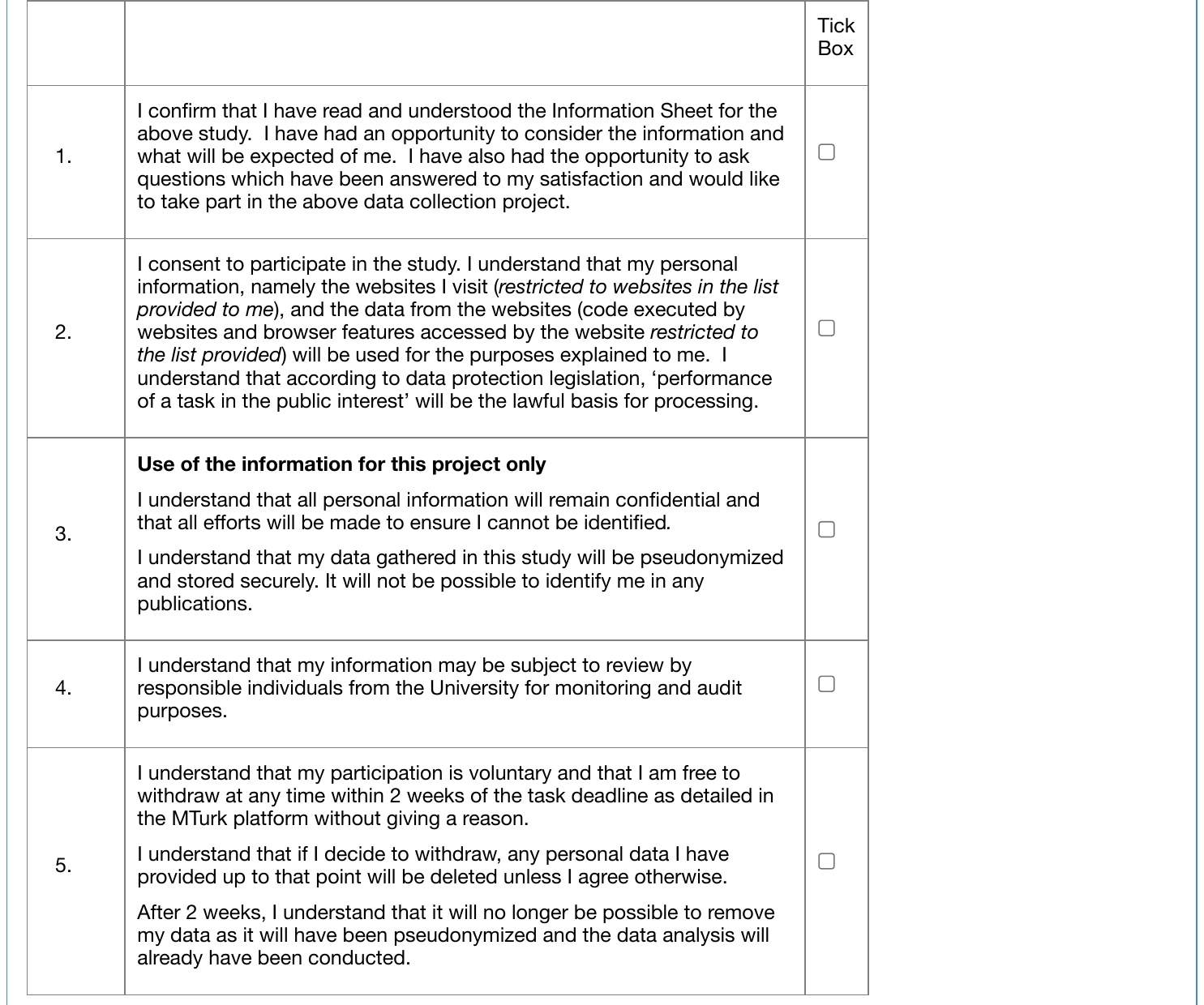}
  \end{subfigure}\\[-1ex]
  \begin{subfigure}{0.5\textwidth}
      \includegraphics[width=\textwidth]{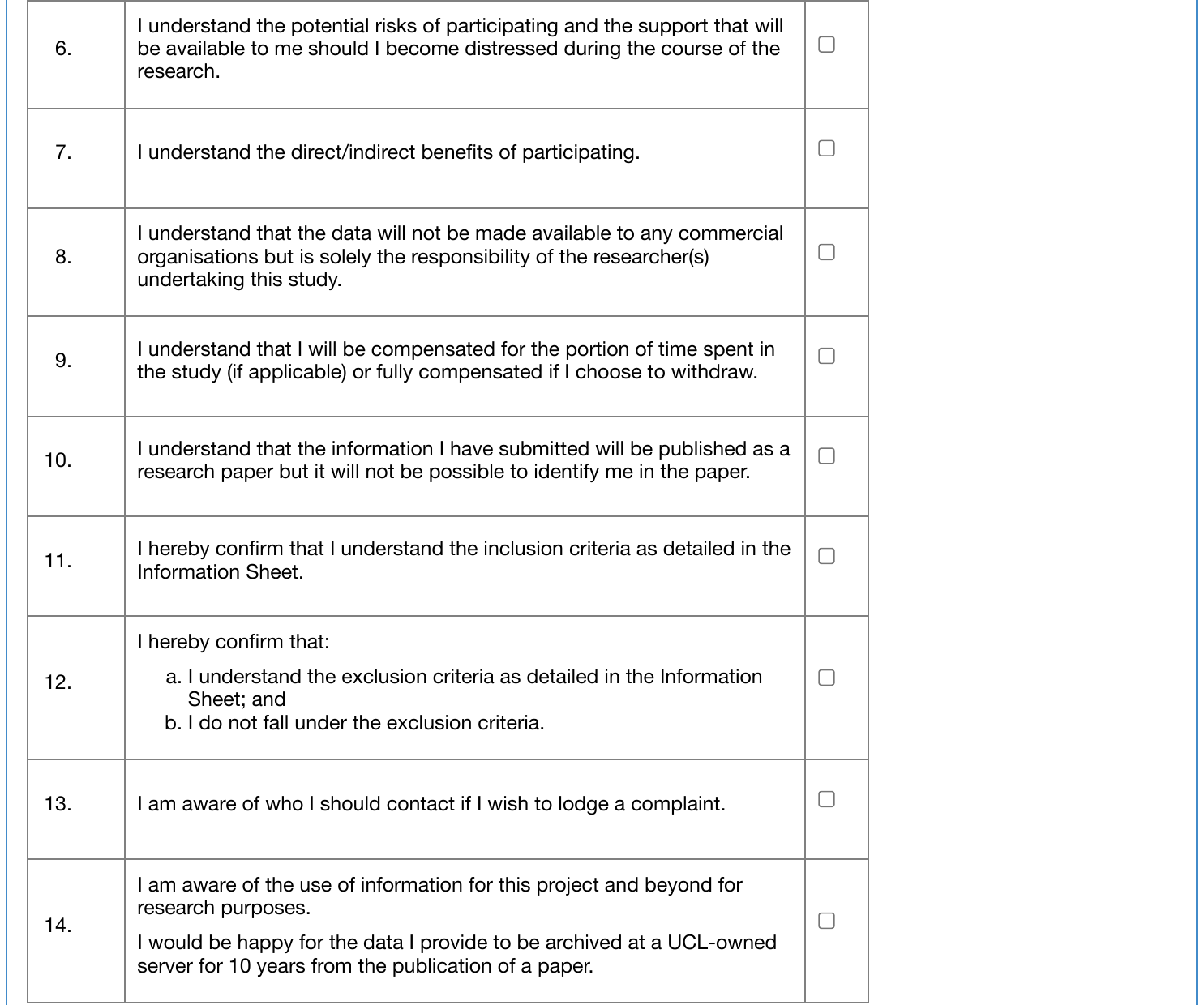}
  \end{subfigure}\\[-1ex]
  \begin{subfigure}{0.5\textwidth}
      \includegraphics[width=\textwidth]{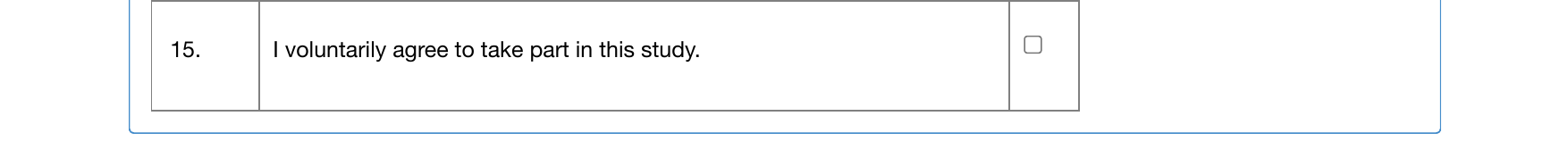}
  \end{subfigure}\\[-0.2ex]
  \begin{subfigure}{0.5\textwidth}
      \includegraphics[width=\textwidth]{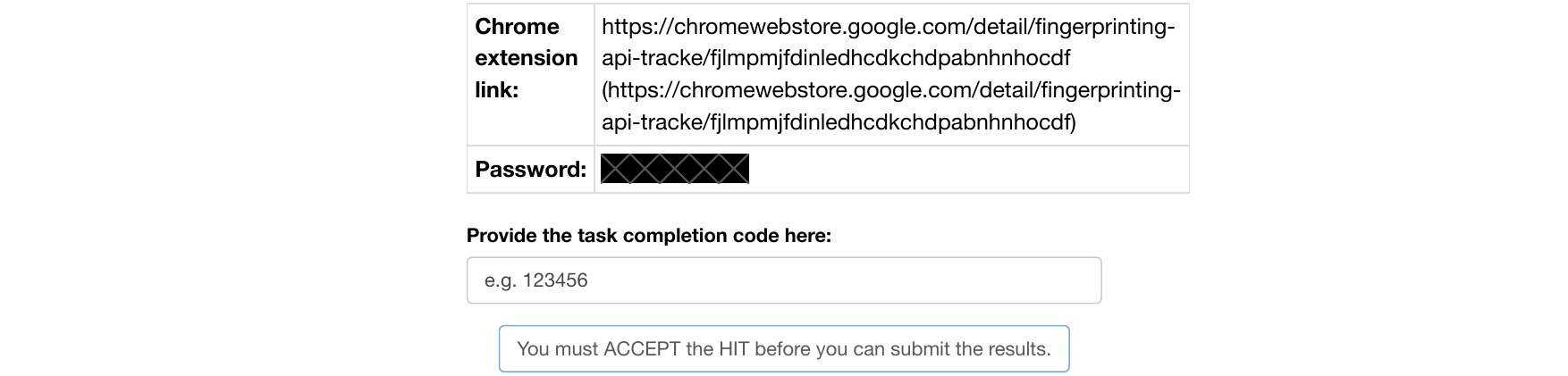}
  \end{subfigure}
  \caption{HIT uploaded to MTurk. To ease readability, a significant portion of the consent form has been left out.}
  \label{fig:hit}
\end{figure*}

In Figure~\ref{fig:hit}, we display the Human Intelligence Task (HIT), i.e., the ad, used to recruit participants from the Amazon mTurk platform.
Specifically, prospective participants were provided with some task instructions, an information sheet describing in layman terms the purpose of the data collection and the types of data being collected and a consent form to ensure they have read and understood their data rights and task requirements.
They were given a password to authenticate themselves with the Chrome extension used for data collection and were told to enter the ``Task Completion Code'', which was provided to them by the extension once they completed browsing the assigned websites.

\end{document}